\documentclass[iicol,pdflatex,sn-mathphys-num]{sn-jnl}
\usepackage{hyperref}
\usepackage{graphicx}
\usepackage{multirow}
\usepackage{amsmath,amssymb,amsfonts}
\usepackage{amsthm}
\usepackage[title]{appendix}
\usepackage{xcolor}
\usepackage{textcomp}
\usepackage{booktabs}
\usepackage{algorithm}
\usepackage{algorithmicx}
\usepackage{algpseudocode}
\usepackage{listings}
\usepackage{lineno}
\usepackage{tabularx}
\usepackage{url}
\usepackage{multirow}
\usepackage{hhline}
\usepackage{subfigure}
\usepackage{xspace}
\DeclareUnicodeCharacter{03B3}{\ensuremath{\gamma}}

\makeatletter
\@ifundefined{@makespecialcolbox}{
  \gdef\@makespecialcolbox{
    \setbox\@outputbox \vbox{
      \@texttop
      \dimen@ \dp\@outputbox
      \unvbox\@outputbox
      \vskip-\dimen@
    }
    \@tempdima \@colht
    \ifdim \wd\@kludgeins>\z@
      \advance \@tempdima -\ht\@outputbox
      \advance \@tempdima \pageshrink
      \setbox\@outputbox \vbox to \@colht{
        \unvbox\@outputbox
        \vskip \@tempdima
        \@textbottom
      }
    \else
      \advance \@tempdima -\ht\@kludgeins
      \setbox\@outputbox \vbox to \@colht{
        \vbox to \@tempdima{
          \unvbox\@outputbox
          \@textbottom
        }
        \vss
      }
    \fi
    {\setbox\@tempboxa \box\@kludgeins}
  }
}{}
\makeatother

\newcommand{\orcidicon}{
  \includegraphics[width=9pt]{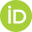}
}
\newcommand{\myorcid}[1]{\href{https://orcid.org/#1}{\orcidicon}}

\newcommand{\ttbar}{\ensuremath{\text{t}\overline{\text{t}}}\xspace}

\begin{document}

\title{Sensitivity of the FCC-ee to axion-like particles at different center-of-mass energies}

\author[1]{Juliette Alimena\myorcid{0000-0001-6030-3191}}
\author[1,2,3]{Elnura Bakhishova\myorcid{0009-0006-0693-7373}}
\author[1,2]{Freya Blekman\myorcid{0000-0002-7366-7098}}
\author[1,4]{Jannah Darwish Abdelhafiz\myorcid{0009-0002-7279-8675}}
\author[1,5]{Christina Dorofeev\myorcid{0009-0008-4321-7273}}
\author[1,6]{Jeremi Niedziela\myorcid{0000-0002-9514-0799}}
\author[7]{Giacomo Polesello\myorcid{0000-0001-8636-0186}}
\author*[1,2]{Anna Przybyl\myorcid{0009-0008-9726-5684}}\email{anna.przybyl@desy.de}
\author[1,2]{Lovisa Rygaard\myorcid{0000-0003-3192-1622}}

\affil[1]{Deutsches Elektronen-Synchrotron DESY, Notkestr. 85, 22607 Hamburg, Germany}
\affil[2]{Institut f\"ur Experimentalphysik, Universit\"at Hamburg, Luruper Chaussee 149, 22761 Hamburg, Germany}
\affil[3]{Department of Physics and Astronomy, University of Amsterdam, 1098 XH Amsterdam, The Netherlands}
\affil[4]{Department of Mathematics, Faculty of Science, Cairo University, Giza 12613, Egypt}
\affil[5]{Institute of Fundamental Science, University of Oregon, 1585 E 13th Ave, Eugene, Oregon 97403, USA}
\affil[6]{Interuniversity Institute for High Energies (IIHE), Vrije Universiteit Brussel, Pleinlaan 2, 1050 Brussels, Belgium}
\affil[7]{INFN Sezione di Pavia, Via Agostino Bassi, 6, 27100 Pavia PV, Italy}

\abstract{The sensitivity of the proposed FCC-ee collider to axion-like particles (ALPs) is investigated at all planned center-of-mass energies, focusing on the case where the ALP couples primarily to electroweak gauge bosons at leading order. We study the associated production of an ALP with a photon, with the ALP decaying in turn to two photons, yielding a three-photon final state. Expected 95\% confidence level sensitivities to the ALP-photon coupling are computed for ALP masses between 5 and 320 GeV. In the benchmark model considered, the FCC-ee can probe the ALP-photon coupling down to a few $10^{-6}$ GeV$^{-1}$ at the Z pole and about $10^{-5}$ GeV$^{-1}$ at the WW, ZH, and \ttbar running stages. The higher-energy running stages extend the mass reach beyond the Z-pole kinematic range, while the Z-pole run provides the best sensitivity at low masses. We also study how the sensitivity changes when the relative coupling strengths to the electroweak gauge fields are varied, showing that the three-photon channel can provide information complementary to photon-fusion probes of ALPs.}

\maketitle

\section{Introduction}

Although widely successful, the Standard Model (SM) of particle physics fails to explain certain experimental observations, such as the baryon asymmetry of the universe, dark matter, and non-zero neutrino masses. Additionally, certain theoretical aspects of the SM are not yet explained, such as the hierarchy problem, the strong CP problem, and flavor structure. These open questions can be addressed by direct observations of new physics (NP) or by measuring deviations from SM predictions, and call for an intensified collider exploration. A high-luminosity, energy frontier lepton collider would allow for precision studies of the electroweak sector. The preferred option for the next generation of high-energy particle colliders is the Future Circular Collider (FCC) at CERN~\cite{fcc_preferred}. The electron-positron stage of the FCC, FCC-ee, would offer numerous opportunities for precision measurements and new-physics studies~\cite{fccee_the_lepton_collider}.

The FCC-ee is proposed to run at or around four different center-of-mass energies ($E_{CM}$): the Z pole (91 GeV), the WW threshold (160 GeV), the ZH production peak (240 GeV), as well as the \ttbar threshold (340-365 GeV) \cite{fcc_vol1, fcc_vol2}. The collision environment will be clean and well-defined, which will allow for precision measurements that are sensitive to tiny deviations from the SM. The high integrated luminosity makes the FCC-ee an ideal facility for direct searches for NP, particularly for processes with small production cross sections or weak couplings that would otherwise be inaccessible. Among the many new-physics scenarios accessible at FCC-ee, ALPs provide a useful benchmark because their collider phenomenology changes strongly with mass, lifetime, and electroweak gauge-boson couplings. Depending on the parameters, ALPs may appear as long-lived particles or as promptly decaying resonances; the present study focuses on the prompt $a\to\gamma\gamma$ regime.

The QCD axion, a pseudo-Nambu-Goldstone boson, was introduced in the 1970s to address the strong CP problem~\cite{Peccei:1977hh,Peccei:1977ur,Weinberg:1977ma,Wilczek:1977pj}. More generally, a broader class of pseudoscalar states, known as ALPs ($a$), occur in theories with spontaneously broken global symmetries, with ALP masses ($m_a$) and couplings to SM particles spanning many orders of magnitude~\cite{collider_probes_of_alps,alps_at_future_colliders}. ALPs are particularly interesting because they provide a generic, but well-motivated, probe of NP across a wide range of energy scales. In certain regions of parameter space, ALPs can be dark matter candidates~\cite{Preskill:1982cy,Abbott:1982af,Dine:1982ah,snowmass_alps_as_dm}, and in others, they can be dark matter mediators~\cite{alps_as_portal,electroweak_axion_portal_to_dark_matter}. At the FCC-ee, the high luminosity at the Z pole run will allow us to probe small ALP couplings, while the higher center-of-mass energies ($E_{CM}$) will, in turn, allow us to probe larger $m_a$.

In the MeV-TeV mass range, ALPs are constrained by collider searches~\cite{currentlimits_faser,currentlimits_besiii_diphoton,currentlimits_besiii,current_limits_atlas_pbpb,current_limits_atlas_pp,currentlimits_cms_pbpb,CMS:2024vjn,TOTEM:2023ewz,CidVidal:2018blh,LHCb:2025gbn}, beam-dump and fixed-target experiments~\cite{currentlimits_na64_beamdump,currentlimits_proton_beamdump,currentlimits_na62_beamdump,currentlimits_slac_beamdump,currentlimits_cern_beamdump,currentlimits_miniboone}, astrophysical bounds from supernovae~\cite{currentlimits_supernovae}, and ALP-specific exclusion reinterpretations~\cite{dEnterria:2021ljz}. Collider searches can be split into two categories: searches via light-by-light scattering, performed at the ATLAS and CMS experiments,~\cite{current_limits_atlas_pbpb,currentlimits_cms_pbpb}, and searches where ALPs decay to photonic final states, performed at the FASER, BESIII, ATLAS, CMS, and LHCb experiments~\cite{currentlimits_faser,currentlimits_besiii,currentlimits_besiii_diphoton,current_limits_atlas_pp,CMS:2024vjn,TOTEM:2023ewz,CidVidal:2018blh,LHCb:2025gbn}. Exclusions at higher masses are presented in Ref.~\cite{dEnterria:2021ljz}, in which exclusions from di-photon searches are reinterpreted in an ALP scenario~\cite{CMS:2011bsw,CMS:2011uvc,ATLAS:2011ab}. A combination of all current exclusions, as well as sensitivity projections for the ALP coupling to photons for Belle II, SHiP, ALICE, MuC, CLIC, ILC, and the FCC, can be found in Ref.~\cite{physics_briefing_book}. In particular, sensitivity projections have been set for the FCC-ee in Refs.~\cite{polesello2025sensitivityfcceedecayaxionlike,photon_fusion} for direct searches, and in Ref.~\cite{schulthess2025newphysicssearchoptical} for beam dump experiments. An overview of ALP signatures at the LHC can be found in Ref.~\cite{alps_at_colliders}, while ALP signatures at future colliders CLIC, CEPC, FCC, SPPC, and MATHUSLA are discussed in Ref.~\cite{alps_at_future_colliders}. In Ref.~\cite{Liu:2022tqn}, ALPs are discussed as an explanation of the $(g-2)_\mu$ anomaly, with a focus on CEPC and FCC-ee, and in Ref.~\cite{Bhattacharya:2025hme}, invisible ALP decays at the ILC are studied.

In this study, we investigate ALPs produced in association with a photon, $e^+ e^- \to a \gamma$, and decaying to two photons, $a \to \gamma \gamma$, at all currently proposed FCC-ee center-of-mass energies. This three-photon final state was previously studied at the Z pole in a more involved analysis~\cite{polesello2025sensitivityfcceedecayaxionlike}, where sensitivity projections were set for $m_a$ ranging from 0.1 to 85 GeV. Here, we update the Z pole study and extend the analysis to the WW, ZH, and \ttbar running stages, thereby probing ALP masses up to 320 GeV. We also study how the reach depends on the relative strengths of the ALP interactions with the electroweak gauge fields. This makes it possible to assess the complementarity of associated ALP production with photon-fusion production in probing the electroweak structure of ALP couplings.

\section{Theoretical framework}

\subsection{Axion-like particles}

The ALP is a pseudoscalar that appears in many SM extensions. The effective Lagrangian of the $a$ coupling to a photon or $Z$ boson after electroweak symmetry breaking can be defined as~\cite{alps_at_future_colliders}
\begin{equation}
\begin{aligned}
\mathcal{L}_{\text{eff}} \ni\;&
e^2 C_{\gamma\gamma}\frac{a}{\Lambda}F_{\mu\nu}\tilde F^{\mu\nu}
+\frac{2e^2}{s_wc_w}C_{\gamma Z}\frac{a}{\Lambda}
F_{\mu\nu}\tilde Z^{\mu\nu} \\
&+\frac{e^2}{s_w^2c_w^2}C_{ZZ}\frac{a}{\Lambda}
Z_{\mu\nu}\tilde Z^{\mu\nu},
\end{aligned}
\label{lagrangian}
\end{equation}
where $c_w = \cos{\theta_w}$ and $s_w = \sin{\theta_w}$. $\Lambda$ is the scale of new physics, $F_{\mu\nu}$ describes the photon in the broken phase of electroweak symmetry, and $Z_{\mu\nu}$ describes the $Z$ boson in the broken phase of electroweak symmetry. The Wilson coefficients $C_{\gamma\gamma}$, $C_{\gamma Z}$, and $C_{ZZ}$ satisfy
\begin{equation}
\begin{aligned}
C_{\gamma\gamma} &= C_{WW} + C_{BB}, \\
C_{\gamma Z} &= c_w^2 C_{WW} - s_w^2 C_{BB}, \\
C_{ZZ} &= c_w^4 C_{WW} + s_w^4 C_{BB},
\end{aligned}
\label{coefficient_relations}
\end{equation}
where $C_{WW}$ is the Wilson coefficient for the ALP coupling to the unbroken $SU(2)$ gauge field, and $C_{BB}$ is the Wilson coefficient for the ALP coupling to the unbroken $U(1)$ gauge field.

\subsection{ALP model benchmark}

\begin{figure*}[htbp]
    \centering
    \subfigure[]{\includegraphics[width=0.45\textwidth]{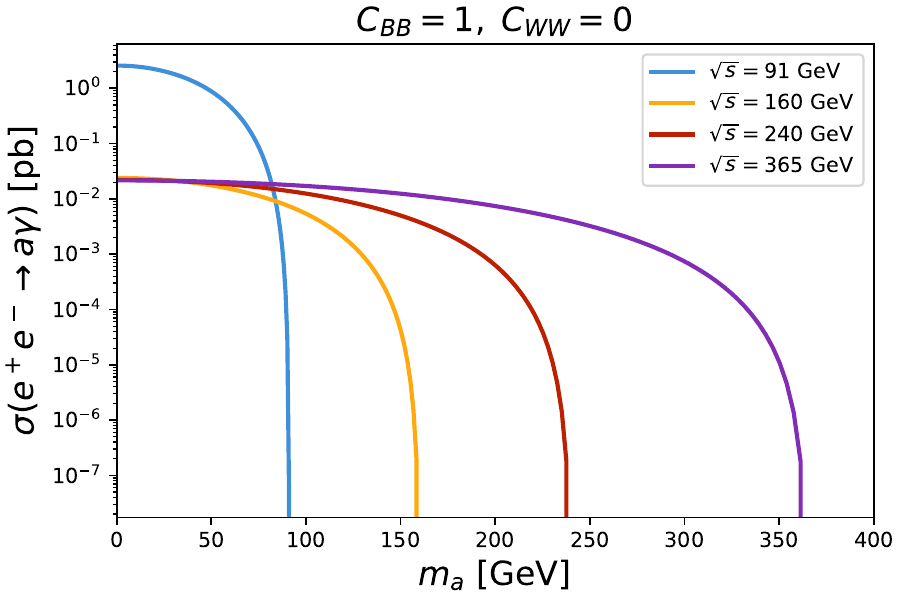}} \qquad
    \subfigure[]{\includegraphics[width=0.45\textwidth]{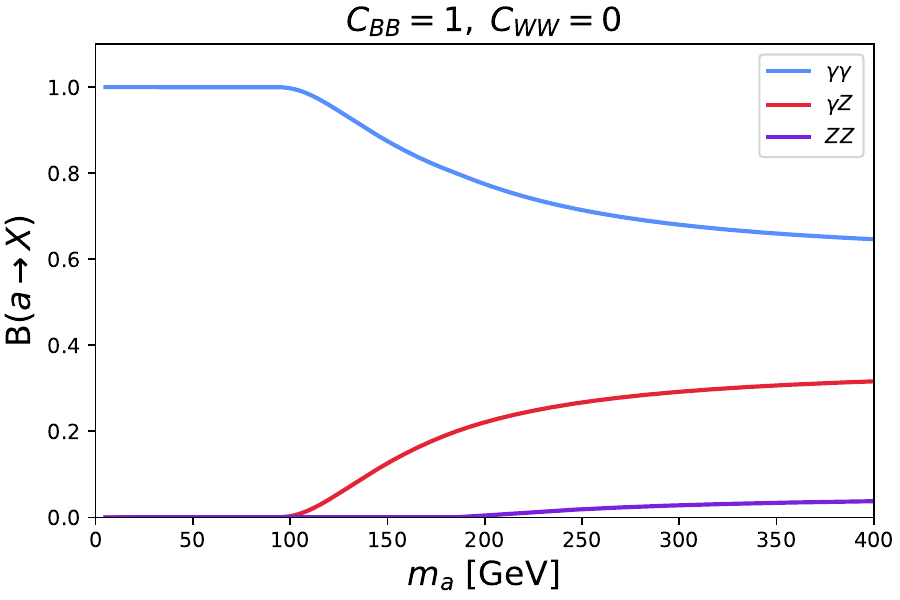}}
    \caption{The production cross section of an ALP produced in association with a photon (a), and the branching ratios of possible ALP decay modes (b). The cross sections are in pb for all proposed runs: Z pole (blue), WW (yellow), ZH (red), and \ttbar running stages (purple). Possible decay modes in our chosen benchmark scenario are $a\to \gamma\gamma$ (blue), $a \to \gamma Z$ (red), and $a \to ZZ$ (purple).}
    \label{fig:br_xsections}
\end{figure*}

We adopt a model benchmark~\cite{alps_at_future_colliders, electroweak_axion_portal_to_dark_matter} where the ALP couples primarily to electroweak gauge bosons by setting $C_{WW} = 0$. The Wilson coefficients now satisfy
\begin{equation}
\begin{split}
    C_{\gamma\gamma} &= C_{BB}, \\
    C_{\gamma Z} &= -s_w^2 C_{BB}, \\
    C_{ZZ} &= s_w^4 C_{BB},
\end{split}
\end{equation}
and imply a dominant coupling to photons in the broken electroweak basis, while retaining a nonzero $a\gamma Z$ interaction that contributes to associated production. We focus on the scenario where an ALP is produced in association with a photon, and the ALP itself decays into two photons. The differential production cross section for an ALP produced in association with a photon is given by~\cite{alps_at_future_colliders}
\begin{equation}
\begin{split}
    \frac{d\sigma(e^+ e^- \to a \gamma)}{d\Omega} = 2 \pi \alpha \alpha^2(s) \frac{s^2}{\Lambda^2} \left( 1 - \frac{m_a^2}{s} \right)^3 \\ \times \left( 1 + \cos^2{\theta} \right) \left( |V_\lambda(s)|^2 + |A_\lambda(s)|^2 \right),
    \label{xsections}
\end{split}
\end{equation}
where $m_a$ is the ALP mass, $\theta$ is the scattering angle of the photon relative to the beam axis, and $\sqrt{s}$ is the center-of-mass energy. The vector and axial-vector form factors are given by
\begin{equation}
\begin{aligned}
V_\lambda(s) &=
\frac{C_{\gamma\gamma}}{s}
+\frac{g_V}{2c_w^2s_w^2}
\frac{C_{\gamma Z}}
{s-m_Z^2+i m_Z\Gamma_Z},
\\[2mm]
A_\lambda(s) &=
\frac{g_A}{2c_w^2s_w^2}
\frac{C_{\gamma Z}}
{s-m_Z^2+i m_Z\Gamma_Z},
\end{aligned}
\label{VA_definitions}
\end{equation}
where $g_V = 2 s_w^2 - 1/2$, $g_A = -1/2$ and $\Gamma_Z$ is the total width of the Z boson. The $\left(1-\frac{m_a^2}{s}\right)^3$ factor leads to a rapid decrease in production as $m_a$ approaches $\sqrt{s}$. Away from the Z pole and for $m_a \ll \sqrt{s}$, the explicit $s^2$ dependence in the cross section is approximately canceled by the $1/s^2$ scaling of the squared form factors, so the higher-energy cross sections approach similar values. At $s = m_Z^2$, the contribution proportional to $C_{\gamma Z}^2$ is enhanced by a factor of $(m_Z^2/\Gamma_Z^2) \approx 1336$ relative to its high-energy limit~\cite{snowmass}. The enhancement of the full cross section is smaller because photon exchange and photon--Z interference also contribute away from the Z pole. These features are shown in Figure~\ref{fig:br_xsections} for $C_{BB}=1$ and $C_{WW}=0$.

Among the electroweak gauge-boson decay modes considered here, only $a\to\gamma\gamma$ is kinematically allowed for $m_a < m_Z$. At $m_a \ge m_Z$ and $m_a \ge 2m_Z$, the $a\to \gamma Z$ and $a \to ZZ$ modes become possible. The corresponding decay widths are
\begin{equation}
\begin{aligned}
\Gamma(a\to \gamma\gamma)
&=
\frac{4\pi\alpha^2 m_a^3 |C_{\gamma\gamma}|^2}
{\Lambda^2},
\\[1ex]
\Gamma(a\to \gamma Z)
&=
\frac{8\pi\alpha^2 m_a^3 |C_{\gamma Z}|^2}
{s_w^2 c_w^2 \Lambda^2}
\left(1-\frac{m_Z^2}{m_a^2}\right)^3,
\\[1ex]
\Gamma(a\to ZZ)
&=
\frac{4\pi\alpha^2 m_a^3 |C_{ZZ}|^2}
{s_w^4 c_w^4 \Lambda^2}
\left(1-\frac{4m_Z^2}{m_a^2}\right)^{3/2},
\end{aligned}
\label{decay_widths}
\end{equation}
and the total decay width is $\Gamma_{\text{tot}} = \Gamma(a\to \gamma\gamma) + \Gamma(a\to \gamma Z) + \Gamma(a\to ZZ)$~\cite{electroweak_axion_portal_to_dark_matter}. The branching ratios ($B$) for the possible decay modes in our effective benchmark are shown in Figure~\ref{fig:br_xsections} for $C_{BB}=1$ and $C_{WW}=0$. For the entire mass range considered in the primary result of this study, the $a\to \gamma\gamma$ decay mode is most probable. In the region of masses/couplings addressed by this analysis, the ALP is dominantly prompt.

\section{Simulation}

\begin{figure*}[ht!]
    \centering
    {\includegraphics[height=0.3\textwidth]{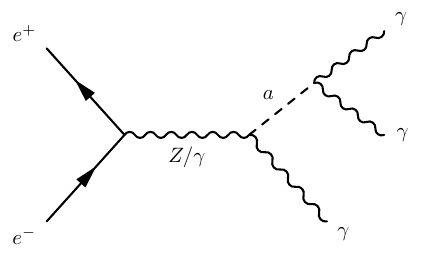}} 
    \hspace{1em}
    {\includegraphics[height=0.3\textwidth]{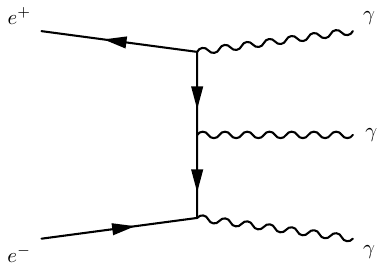}} 
    \caption{The diagrams for the signal process (left) and background process (right).}
    \label{fig:diagrams}
\end{figure*}

The signal process that is simulated is
\begin{equation}
    e^+ e^- \to a\gamma, \qquad a\to \gamma\gamma, 
\end{equation}
and the Feynman diagram for this process is shown in Figure~\ref{fig:diagrams}. The most common irreducible background process that yields a $3\gamma$ final state is
\begin{equation}
    e^+ e^- \to \gamma\gamma\gamma,
\end{equation}
the Feynman diagram for which can also be seen in Figure~\ref{fig:diagrams}. Only the irreducible $e^+e^- \to \gamma\gamma\gamma$ background is simulated explicitly. Reducible backgrounds from electron or neutral-hadron misidentification, photon conversions, and beam-related effects are detector-dependent and are not modeled in this Delphes-based study.

\subsection{Event simulation}

Events are generated for the signal using the Monte Carlo event generator \texttt{MG5aMC@NLO} v. 3.6.3~\cite{Madgraph} along with the \texttt{ALP UFO} v. 1.1~\cite{alps_at_future_colliders} model. All couplings are set to 0 except for $C_{BB}$, which is set to 1 and yields
\begin{equation}
    C_{\gamma \gamma} = 1, \qquad C_{\gamma Z} = -0.223.
\end{equation}
The $\Lambda$ parameter is set to 1 TeV, and 1 million events are generated for each $m_a$ point. The generated ALP mass ranges can be found in Table~\ref{ALP_mass_ranges}, with the exact mass points for each running stage in Table~\ref{tab:cross_sections} in the Appendix. Photons are required to have a minimum energy $E_\gamma > 0.1$ GeV and a pseudorapidity $|\eta| <  2.6$ at the generation level. The pseudorapidity requirement matches the reconstructed-photon acceptance of the Delphes IDEA card, as described below, and removes the very forward region.

\begin{table*}[ht]
\begin{center}
\begin{tabular}{ c | c | c | c }
Running stage & $E_{CM}$ [GeV] & $m_a$ range [GeV] & Integrated luminosity [ab$^{-1}$] \\ 
 \hline 
 Z & 91 & 5--80 & 205 \\  
 WW & 160 & 5--150 & 19.2 \\ 
 ZH & 240 & 5--220 & 10.8 \\ 
 \ttbar & 365 & 5--320 & 2.7 \\ 
\end{tabular}
\caption{Information about all runs and simulated $m_a$ ranges, with integrated luminosities for each run taken from Ref.~\cite{fcc_vol1}.}
\label{ALP_mass_ranges}
\end{center}
\end{table*}

The $3\gamma$ background samples are also generated using \texttt{MG5aMC@NLO} v. 3.6.3, using the default SM model input. Approximately 1 million events are generated at each $E_{CM}$.

For both signal and background events, the showering and hadronization are simulated using PYTHIA8~\cite{pythia}, and a fast detector simulation is performed using Delphes~\cite{delphes}. One of the proposed general-purpose detectors to be installed at the FCC-ee is the Innovative Detector for Electron-positron Accelerators (IDEA)~\cite{fcc_idea_concept}. A high-precision vertex detector coupled with a high-granularity calorimeter will lead to high-resolution energy measurements, and the detector will have very good particle identification. In particular, the calorimeter system comprises a crystal electromagnetic calorimeter combined with a dual-readout scintillator/fiber hadronic calorimeter~\cite{detector_requirements}. The official IDEA datacard in the ``Winter2023'' production~\cite{FCC-eeConfig} of the FCC-PED study is used in the Delphes simulation. In the final step, the events are analyzed with the FCCAnalyses v. 0.11.0 framework~\cite{fccanalyses}. Since the current analysis depends on the simulation of the electromagnetic calorimeter, the photon energy resolution is adjusted to reflect the predicted performance specified by the IDEA detector. For an energy $E$ in GeV, the photon resolution is defined as
\begin{equation}
    \sigma(E) = \sqrt{0.005^2 E^2 + 0.03^2 E + 0.002^2}
\end{equation}
in the Delphes IDEA card. Photon reconstruction is performed using the default Delphes IDEA card, so the photons are required to have an energy $E \ge 2$ GeV and a $|\eta|\le 2.6$ at reconstruction level, as described in Section~\ref{sec:preselection}.

\section{Event selection}
\label{m_cut_identification}

\begin{figure}[ht!]
    \centering
    {\includegraphics[width=0.45\textwidth]{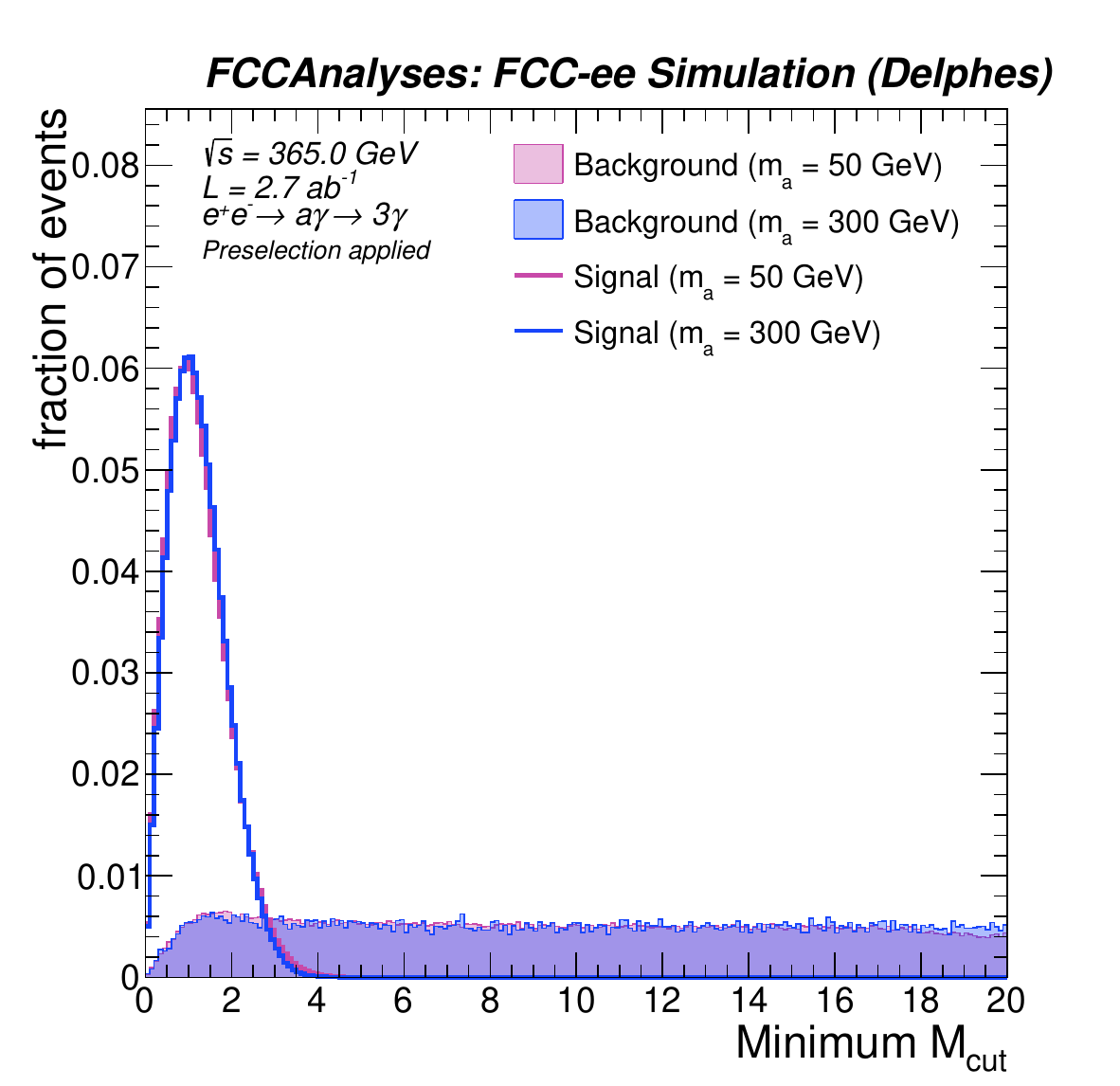}}
    \caption{The $M_{cut}$ distribution after preselection, using the photon assignment that minimizes $M_{cut}$, at the \ttbar threshold for $m_a$ = 50 GeV (pink) and 300 GeV (blue). The solid lines correspond to signal samples, while the shaded regions correspond to background samples. The number of events for each distribution is normalized to unity.}
    \label{fig:m_cut_plot}
\end{figure}

The first step in the analysis is to assign which photons come from the decay of the ALP (denoted $\gamma_1$ and $\gamma_2$, with $E_{\gamma_1} > E_{\gamma_2}$), and which one is produced in association with the ALP (denoted $\gamma_r$). According to the recoil formula, the photon produced in association with the ALP will have an energy of
\begin{equation}
    E_r = \frac{E_{CM}^2 - m_a^2}{2 E_{CM}}.
\end{equation}
We then define a variable \cite{polesello2025sensitivityfcceedecayaxionlike}
\begin{equation}
    M_{cut}^2 = \frac{(m_{\gamma_1 \gamma_2} - m_a)^2}{\sigma(m_a)^2} + \frac{(E_{\gamma_r} - E_r)^2}{\sigma(E_r)^2},
\end{equation}
where $\sigma(E_r)$ is the photon energy resolution, and $\sigma(m_a)$ is the ALP mass resolution. We iterate over all possible assignments of the three photons in the final state and choose the photon assignment that minimizes $M_{cut}^2$. This choice is the best identification given the expected kinematics of the ALP and corresponding recoil photon. The minimum $M_{cut}$ is shown in Figure~\ref{fig:m_cut_plot} for two benchmark points $m_a$ = 50 and 300 GeV at the \ttbar threshold. For the signal, the minimum $M_{cut}^2$ peaks at a low value, while for the background, there is no peak. The signal and background distributions are rather independent of $m_a$, and the behavior is similar across all $E_{CM}$.

Given the same generated background, different photons will be chosen as ALP decay products depending on the ALP mass point. Thus, each ALP mass point yields its own analysis; there are different selection criteria for each mass point, and the overall result is mass-dependent.

\subsection{Preselection}
\label{sec:preselection}

The selections are done in two stages: preselection and final selection. The purpose of the preselection is to clean up the events by choosing only those in which the three photons correspond to the collision, and they can all be resolved. The same preselection is applied to all ALP mass points, while the final selection is optimized separately for each tested ALP mass hypothesis. The effect of the preselection is shown in Table~\ref{seleff_backrej} in the Appendix, which lists signal efficiency and background rejection values for each center-of-mass energy and $m_a$ value. Typically, about 90\% of the signal is kept, and almost 50\% of the $e^+ e^- \to \gamma\gamma\gamma$ background is rejected. This preselection is also expected to suppress potential background processes involving Higgs boson decays to two photons to a negligible level.

The following selections are applied as a preselection:
\begin{itemize}
    \item The number of reconstructed photons is required to be exactly three, in order to remove possible backgrounds. Photons are required to have $E \ge 2$ GeV and $|\eta|\le 2.6$, and are assigned according to the corresponding ALP kinematics, as described above.
    \item The invariant mass of the three photons is required to be near $E_{CM} \pm\sigma(m_{\gamma\gamma\gamma})$. The resolution $\sigma(m_{\gamma\gamma\gamma})$ depends on the $E_{CM}$ and can be calculated from the simulation. The values of $\sigma(m_{\gamma\gamma\gamma})$ used in this study are 1.3 GeV (Z pole run), 2.1 GeV (WW threshold), 3.0 GeV (ZH production peak), and 4.2 GeV (\ttbar threshold). This requirement is meant to remove any instrumental background, in which a particle may have escaped the detector, or a photon was mis-reconstructed.
    \item The 3D opening angle between the ALP decay products ($\Delta \alpha_{\gamma_1 \gamma_2}$) must be larger than 0.02. This selection is applied to make sure that all photons are properly resolved in the detector~\cite{polesello2025sensitivityfcceedecayaxionlike}.
\end{itemize}

\subsection{Final selection}

\begin{figure*}[h]
    \centering
    \subfigure[]{\includegraphics[width=0.45\textwidth]{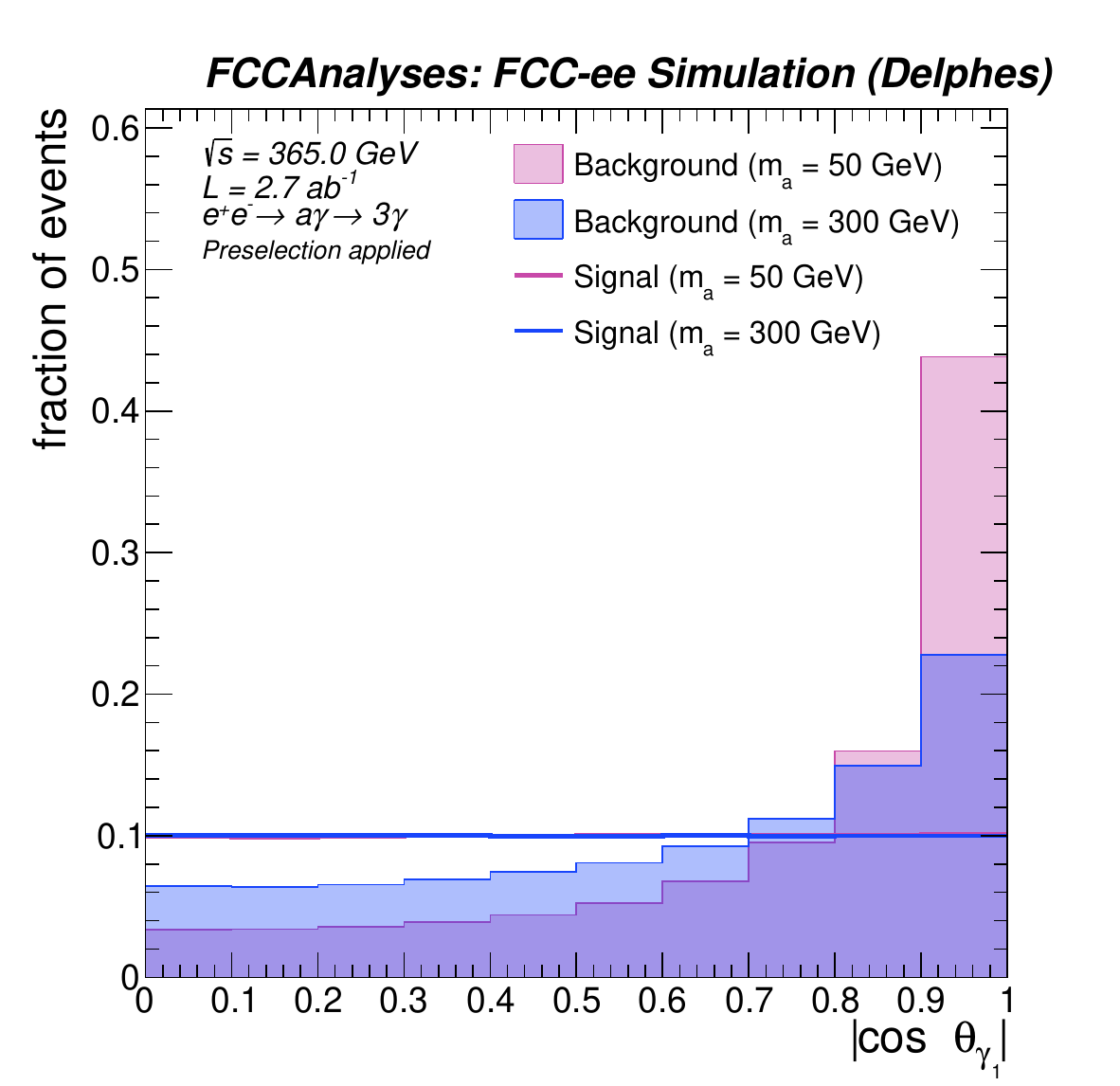}} 
    \subfigure[]{\includegraphics[width=0.45\textwidth]{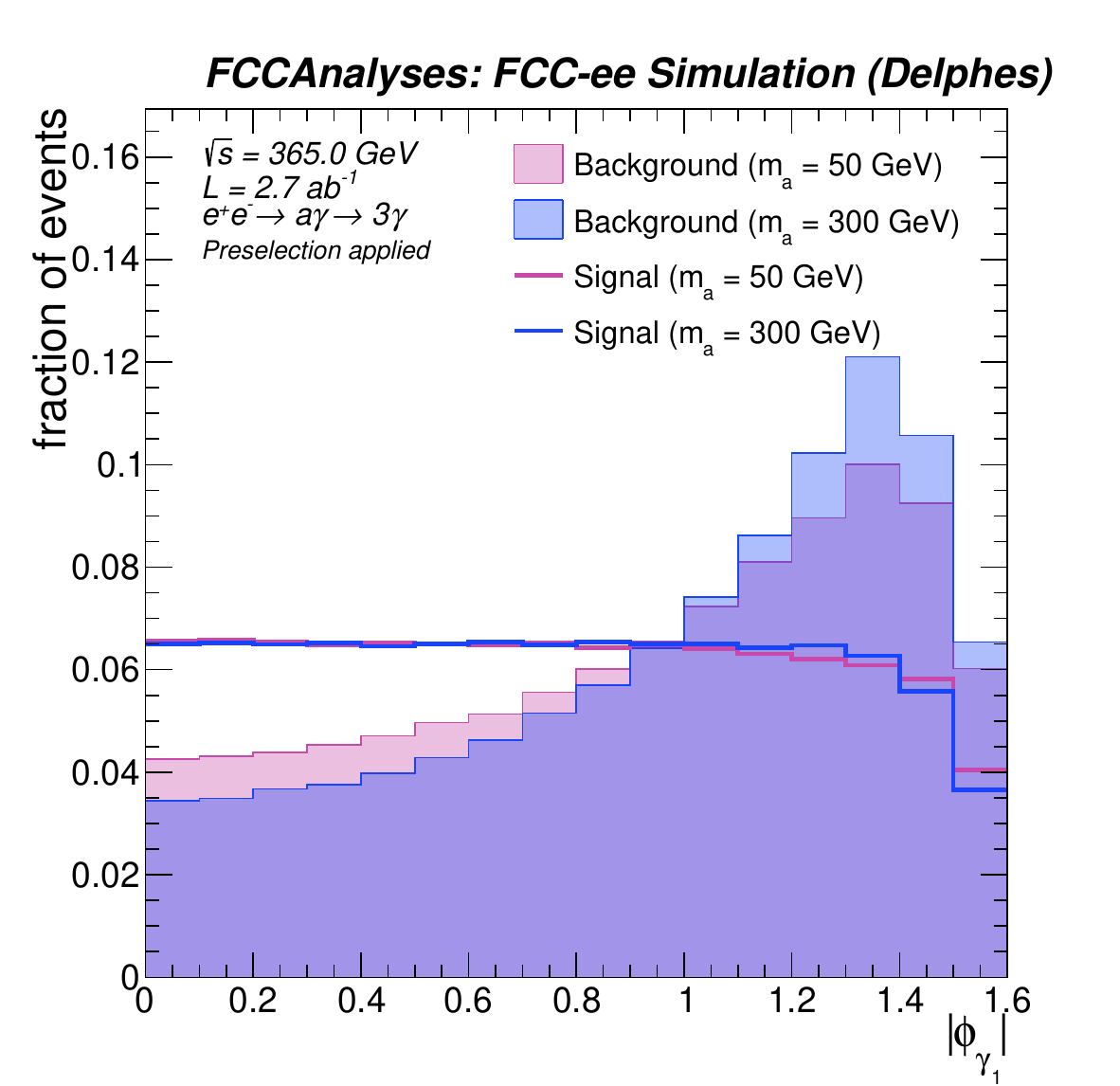}}
    \subfigure[]{\includegraphics[width=0.45\textwidth]{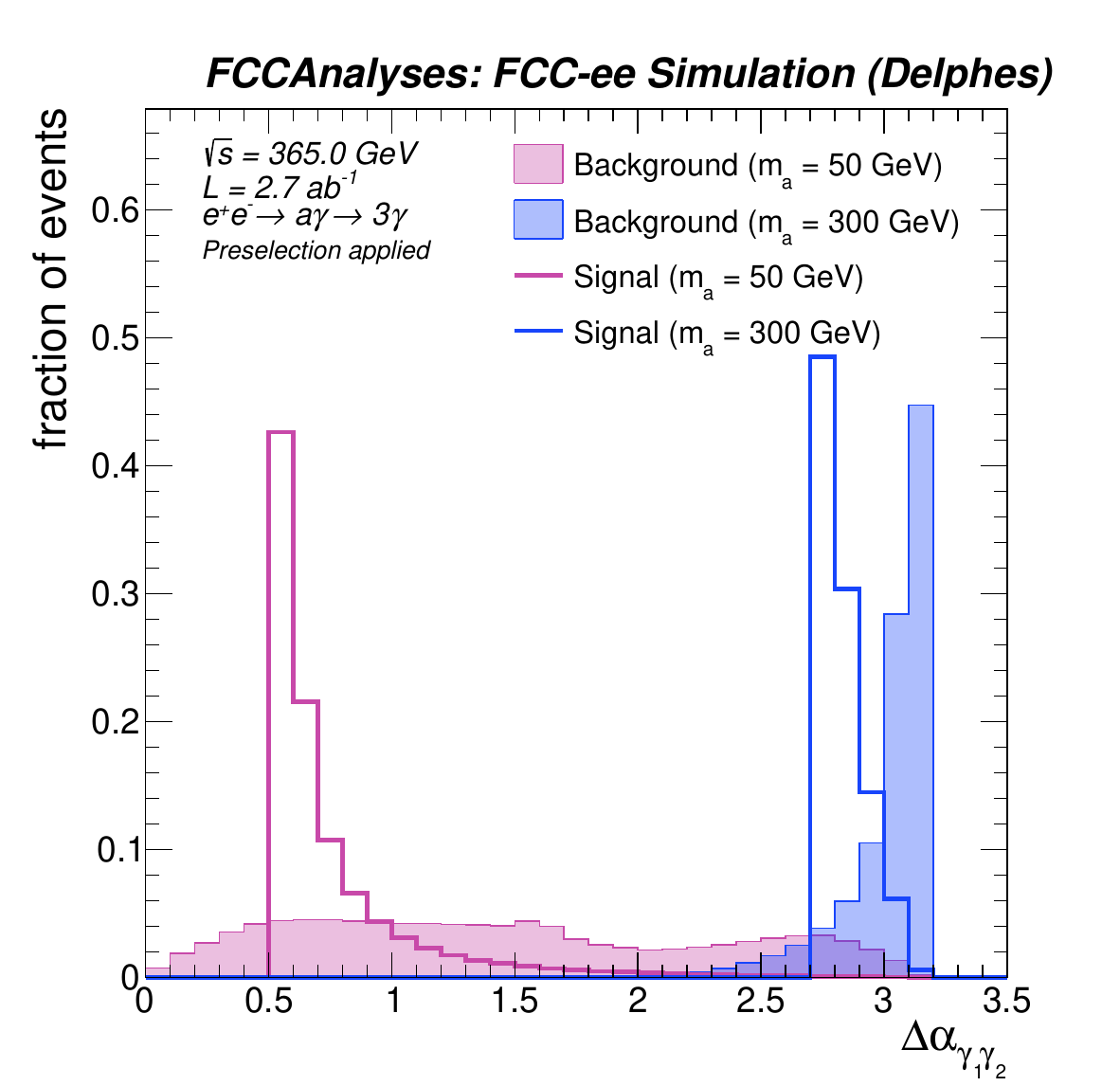}}
    \caption{Three of the discriminant variables used in the final event selection at the \ttbar threshold for $m_a$ = 50 GeV (pink) and 300 GeV (blue), including $|\cos{\theta_{\gamma_1}}|$ (a), $|\phi_{\gamma_1}|$ (b), and $\Delta \alpha_{\gamma_1 \gamma_2}$ (c). The solid lines correspond to signal samples, while the shaded regions correspond to background samples. The preselection is applied to the events, and the number of events for each distribution is normalized to unity.}
    \label{fig:cuts}
\end{figure*}

As shown in Section~\ref{m_cut_identification}, the three final-state photons are assigned according to the expected kinematics of the collision. The signal consists of two-body associated production followed by the two-body ALP decay, which means that each event is fully defined by five variables: the reconstructed ALP mass ($m_{\gamma_1 \gamma_2}$), the $\cos{\theta}$ and $\phi$ of the recoil photon in the lab frame, and the $\cos{\theta}$ and $\phi$ of one of the ALP decay products in the rest frame of the ALP. The system has cylindrical symmetry around the beam axis, so all of the events are rotated such that $\phi_{\gamma_r} = 0$. Then, the following variables are used to select events, the first of which is shown in Figure~\ref{fig:m_cut_plot}, and the rest of which are shown in Figure~\ref{fig:cuts} at the \ttbar threshold:
\begin{itemize}
    \item Minimum $M_{cut}$ (defined in Section \ref{m_cut_identification})
    \item $\cos{\theta_{\gamma_1}}$ (calculated in the ALP rest frame)
    \item $\phi_{\gamma_1}$ (calculated in the ALP rest frame)
    \item $\Delta \alpha_{\gamma_1 \gamma_2}$ (the 3D opening angle between the ALP decay products).
\end{itemize}

A set of final selection criteria is applied to these four variables. These final selection criteria are determined by maximizing the significance as defined in Ref.~\cite{Punzi2003SensitivityOS}. The selection criteria and cutflow tables for all ALP mass points and $E_{CM}$ are provided in the Appendix. The distributions behave in a similar way at the different center-of-mass energies, so only the \ttbar threshold is shown here. For example, a benchmark point of $m_a = 25$ GeV at the Z pole behaves similarly to a benchmark point of $m_a = 100$ GeV at the \ttbar threshold, as both of these points have $m_a/E_{CM} \approx 0.27$.

The event normalization is sample-dependent. For signal at the Z pole run, the event yield is normalized to the expected $6 \times 10^{12}$ Z bosons. This accounts for the FCC-ee baseline running scenario, in which the Z pole dataset is distributed across three center-of-mass energies near the resonance \footnote{Effectively, this is done by first normalizing the number of events to 205 ab$^{-1}$, and then multiplying by a factor of 0.49. This factor corresponds to the expected number of Z bosons produced at an integrated luminosity of 205 ab$^{-1}$, compared to the expected number of Z bosons produced during the entirety of the FCC-ee Z pole run.}. For background at the Z pole, as well as for the signal and background at the other running stages, the event yields are normalized to the respective cross sections and integrated luminosities, given in Tables~\ref{tab:cross_sections} and \ref{ALP_mass_ranges}, respectively.

Since $a$ is a spin-0 particle, $|\cos{\theta_{\gamma_1}}|$ and $|\phi_{\gamma_1}|$ are flat for the signal, while they peak around 1 and $\pi/2$, for the background, respectively. The 3D opening angle $\Delta \alpha_{\gamma_1 \gamma_2}$ depends on $m_a$. At small $m_a$, the ALP is boosted, and the decay products are collimated, so the angle between $\gamma_1$ and $\gamma_2$ is close to 0. As the ALP mass increases, the ALP becomes less boosted, and thus the decay products are less collimated. When $m_a$ is very near to the $E_{CM}$, the angle between $\gamma_1$ and $\gamma_2$ is very close to $\pi$.

The behavior of $|\cos{\theta_{\gamma_1}}|$ and $|\phi_{\gamma_1}|$ is similar across all $m_a$ for the signal, but not for the background, since the photon assignment is based on the desired ALP kinematics. The same reasoning can be applied to the background samples regarding the $\Delta \alpha_{\gamma_1 \gamma_2}$ variable. The signal samples do not behave the same way across all $m_a$ for $\Delta \alpha_{\gamma_1 \gamma_2}$ (see above).

\section{Results}

In Section~\ref{primary_results}, primary results are presented for the $C_{BB}=1$ benchmark scenario. The results are recast to study the relationship between $C_{WW}/C_{BB}$ in Section~\ref{recast_limits}.

\subsection{Primary results}
\label{primary_results}

\begin{figure}[ht!]
    \centering
    {\includegraphics[width=0.45\textwidth]{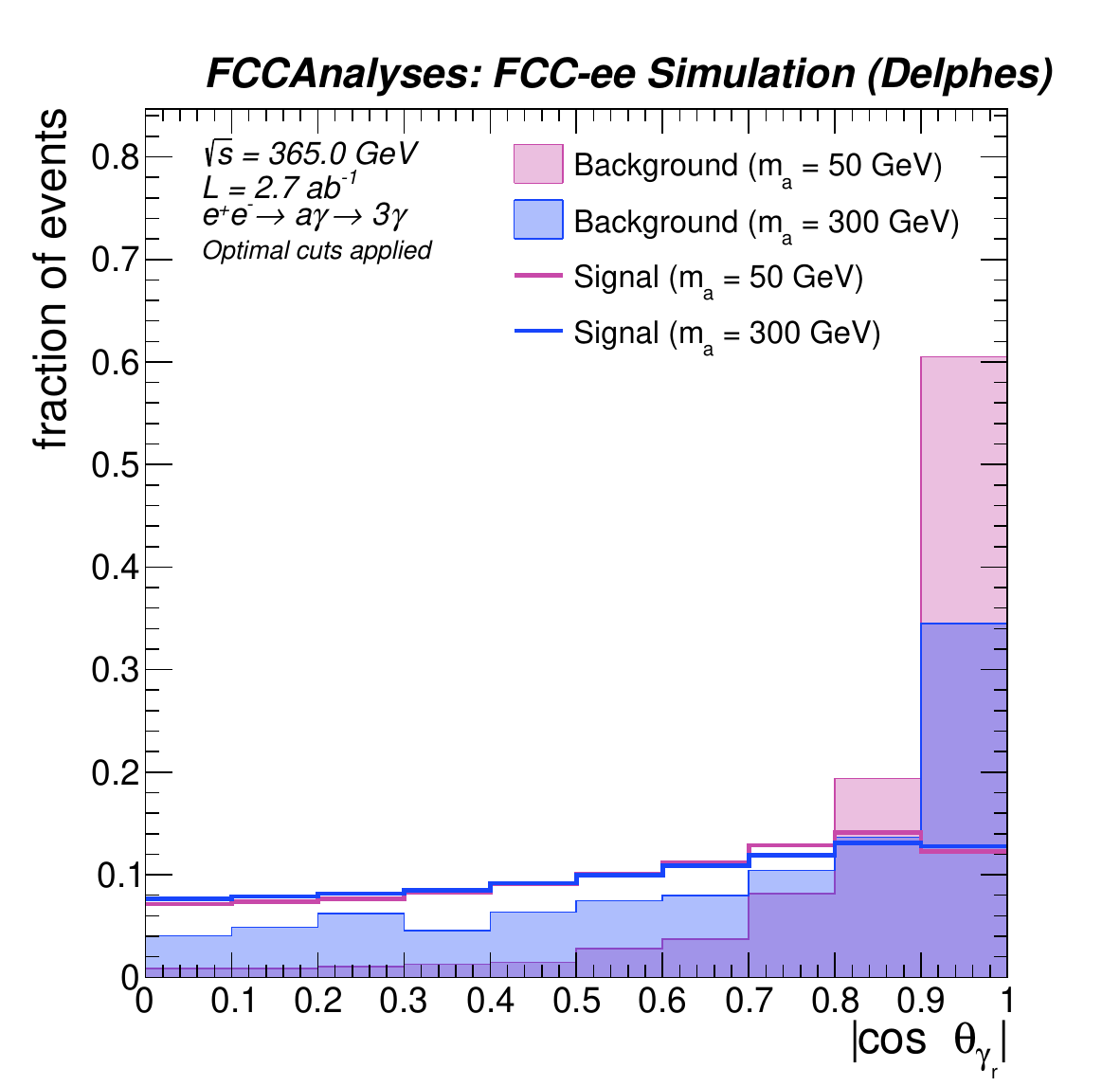}}
    \caption{The $|\cos{\theta_{\gamma_r}}|$ distribution at the \ttbar threshold after final selection for $m_a =$ 50 GeV (pink), and 300 GeV (blue). The solid lines correspond to signal samples, while the shaded regions correspond to background samples. The number of events for each distribution is normalized to unity.}
    \label{fig:cosr_plot}
\end{figure}

The expected number of events for all runs, after the mass-dependent final selection is applied, is shown in Tables~\ref{zpole_selections}-\ref{tt_selections}. The expected sensitivity is then computed separately for each tested ALP mass hypothesis using these event yields. The statistical analysis is performed with the CMS \textsc{Combine} statistical tool~\cite{combine_tool_2024}, based on the \textsc{RooFit}~\cite{roofit} and \textsc{RooStats}~\cite{roostats} frameworks. A binned likelihood fit is carried out using the $|\cos{\theta_{\gamma_r}}|$ distribution as input, as shown in Figure~\ref{fig:cosr_plot} at the \ttbar threshold. The distributions behave in a similar way at the different center-of-mass energies, so we choose to show only the \ttbar threshold case here. The signal distributions are relatively flat, while the background distributions peak around 1. Limits on the expected signal exclusion are derived using the modified frequentist $\text{CL}_\text{s}$ criterion~\cite{CLS1,Junk:1999kv}, with the profile likelihood ratio as the test statistic~\cite{Cowan:2010js}. The asymptotic approximation to the profile likelihood test statistic~\cite{Cowan:2010js} is used. No systematic uncertainties are included in the likelihood. The quoted limits should therefore be interpreted as statistical projections based on the simulated signal and irreducible-background samples.

The output of this procedure is the expected upper limit on the signal strength parameter $\mu=\sigma/\sigma_{theory}$. The corresponding cross section limit satisfies $\mu_{lim} = \sigma_{lim}/\sigma_{theory}$. For the fixed benchmark relation between $C_{\gamma\gamma}$ and $C_{\gamma Z}$, the production cross section and signal yield scale quadratically with the overall coupling normalization. The limit on the coupling is found via
\begin{equation}
    C_{\gamma\gamma, lim} = C_{\gamma\gamma,theory} \sqrt{\frac{\sigma_{lim}}{\sigma_{theory}}},
\end{equation}
where $C_{\gamma\gamma,theory}$ and $\sigma_{theory}$ correspond to our $C_{\gamma\gamma}=1$ benchmark scenario. The limits are shown in terms of $g_{a\gamma\gamma}$, following the current reference reach plots for the ALP-photon coupling~\cite{physics_briefing_book}, and current ALP-photon coupling exclusions set by previous experiments~\cite{AxionLimits}. The conversion from $C_{\gamma\gamma}$ to $g_{a\gamma\gamma}$ is performed with the equation
\begin{equation}
    g_{a\gamma\gamma} = 4 e^2 \cdot \frac{C_{\gamma\gamma}}{\Lambda},
\end{equation}
where $C_{\gamma\gamma} = C_{\gamma\gamma,lim}$, and $\Lambda = 1$ TeV.

\begin{figure*}[htpb]
    \centering
    {\includegraphics[width=0.75\textwidth]{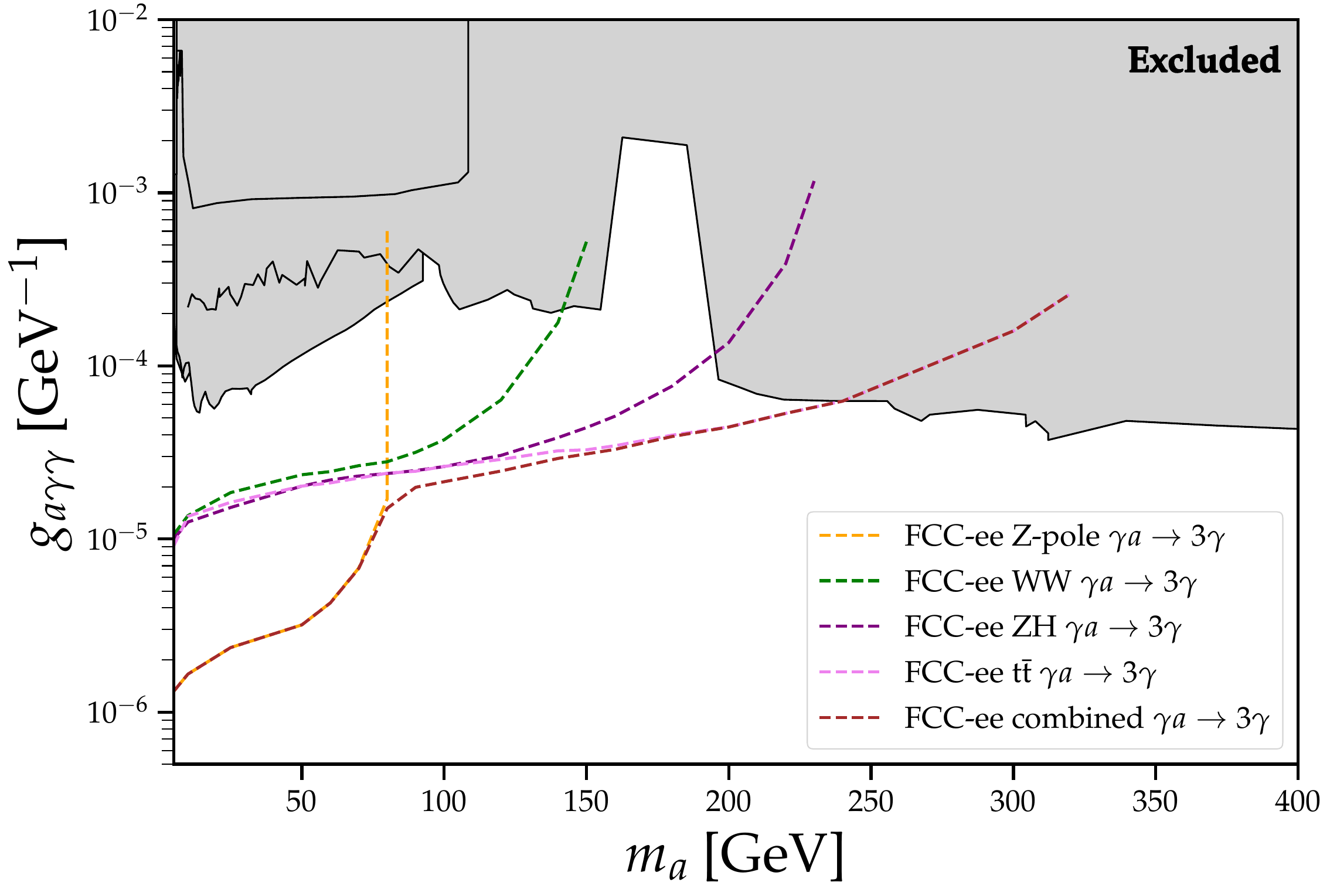}}
    \caption{The 95\% CL projected sensitivity of the FCC-ee, assuming four $e^+ e^-$  interaction points, in the $g_{a\gamma\gamma}$-$m_a$ plane, from the $3\gamma$ final state at the Z pole (yellow), the WW (green), the ZH (purple), the \ttbar running stages (pink), and the combined reach of all proposed runs (brown). The grey regions indicate the areas excluded by previous experiments~\cite{AxionLimits}.}
    \label{fig:limits_closeup}
\end{figure*}

The 95\% confidence level (CL) projected sensitivity of the FCC-ee to $g_{a\gamma\gamma}$ is shown in Figure~\ref{fig:limits_closeup} at each proposed FCC-ee run, as well as the combined sensitivity over all runs. The integrated luminosities correspond to the FCC-ee baseline operation scenario with four interaction points, and therefore represent the full FCC-ee dataset in this configuration. Below ALP masses of 80 GeV, the strongest sensitivity is obtained at the Z pole run, reaching ALP-photon couplings of a few $10^{-6}$ GeV$^{-1}$. The sensitivity at higher-energy running stages is roughly an order of magnitude weaker in this region, but these stages extend the accessible ALP mass range to larger $m_a$. In the 80 to 150 GeV mass range, multiple runs contribute to the sensitivity. Combining data from different collision energies improves the sensitivity to $g_{a\gamma\gamma}$ by up to a factor of 1.7, depending on the mass, relative to each individual run.

\begin{figure*}[htpb]
    \centering
    {\includegraphics[width=0.75\textwidth]{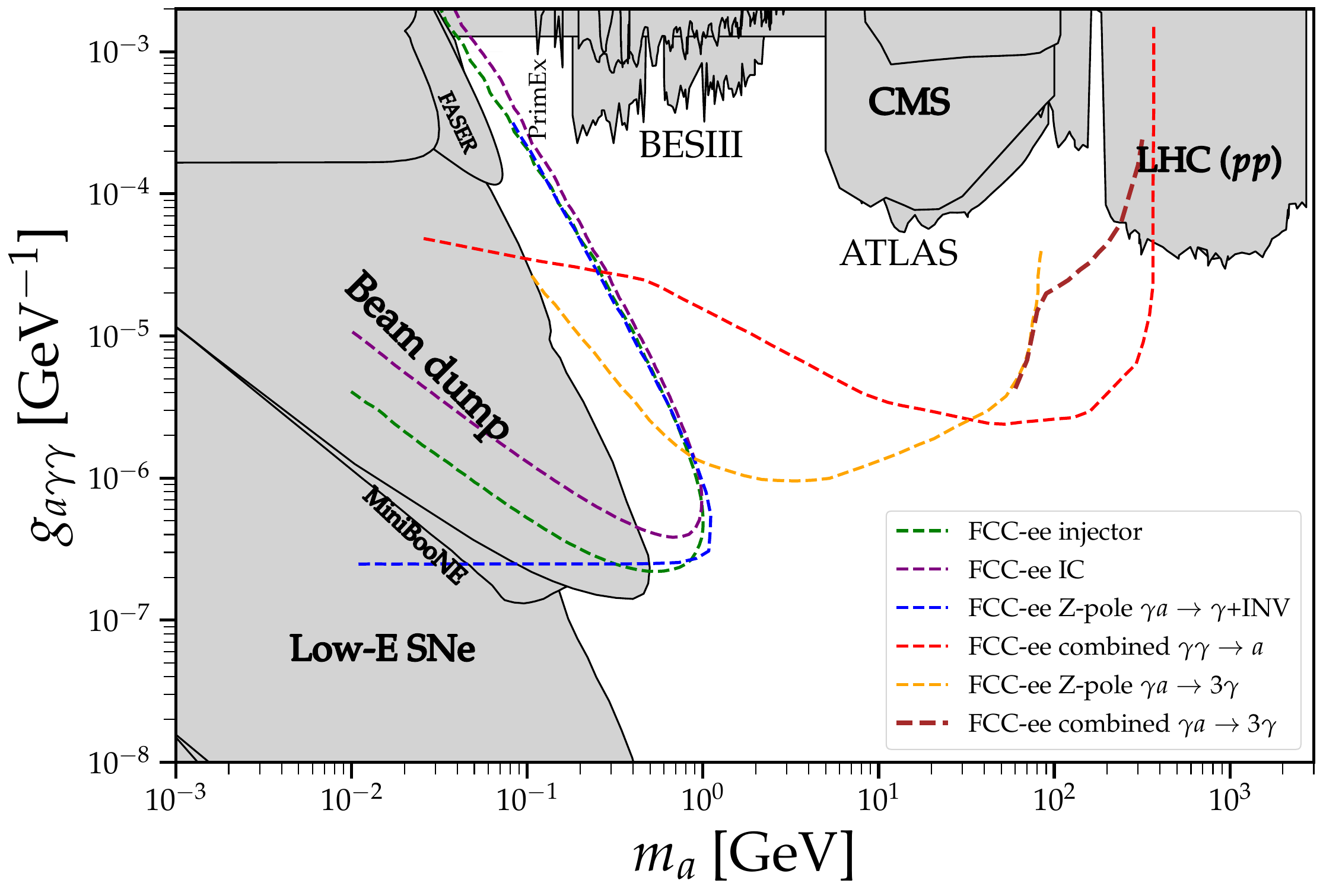}}
    \caption{The 95\% CL projected sensitivity of the FCC-ee, assuming four $e^+ e^-$  interaction points, in the $g_{a\gamma\gamma}$-$m_a$ plane. The brown line indicates the limit for the $3\gamma$ final state from this work, for all center-of-mass energies combined. The blue line indicates the monophoton case, when the ALP decays outside of the detector~\cite{polesello2025sensitivityfcceedecayaxionlike}. The yellow line corresponds to the $3\gamma$ final state at the Z pole, with special attention given to $0.1 < m_a < 5$ GeV~\cite{polesello2025sensitivityfcceedecayaxionlike}. The red line shows the case where the ALP is produced via $\gamma\gamma$-fusion~\cite{photon_fusion}, for all center-of-mass energies combined. The green and purple lines show the case where the ALP is detected via diphoton decays in beam dump experiments at the FCC-ee~\cite{schulthess2025newphysicssearchoptical}. The green line corresponds to the case where the FCC-ee injector complex is used part-time between times for the beyond-collider physics program. The purple line corresponds to the case where the intensity control of the collider applies Compton backscattering to balance the population of the colliding bunches. The grey regions indicate the areas excluded by previous experiments~\cite{AxionLimits}.}
    \label{fig:limits_all}
\end{figure*}

The results from this study can be found in Figure~\ref{fig:limits_all}, along with previous results from Ref.~\cite{polesello2025sensitivityfcceedecayaxionlike,photon_fusion,schulthess2025newphysicssearchoptical}, and already excluded regions of parameter space. In particular, the combined $3\gamma$ result improves upon the current LHC limits for $90 < m_a < 300$ GeV. The $\gamma\gamma$-fusion ALP production mode ($\gamma\gamma \to a \to \gamma\gamma$) results obtained from Ref.~\cite{photon_fusion} have better sensitivity in this mass region, and are combined for all proposed center-of-mass energies. However, these results are sensitive only to $C_{\gamma\gamma}$, and assume $B(a\to \gamma\gamma) = 1$. In contrast, the $3\gamma$ final state is sensitive to both $C_{\gamma\gamma}$ and $C_{\gamma Z}$, and the primary benchmark includes the additional $a \to \gamma Z$ and $a \to ZZ$ decay modes. In the case of a discovery, measurements of ALP production in both the $\gamma\gamma$-fusion and $3\gamma$ channels would help constrain the relative ALP couplings to the $SU(2)$ and $U(1)$ gauge fields.

\subsection{Dependence of the results on the Wilson coefficients in the Lagrangian}
\label{recast_limits}

\begin{figure*}[ht!]
    \centering
    \subfigure[]{\includegraphics[width=0.45\textwidth]{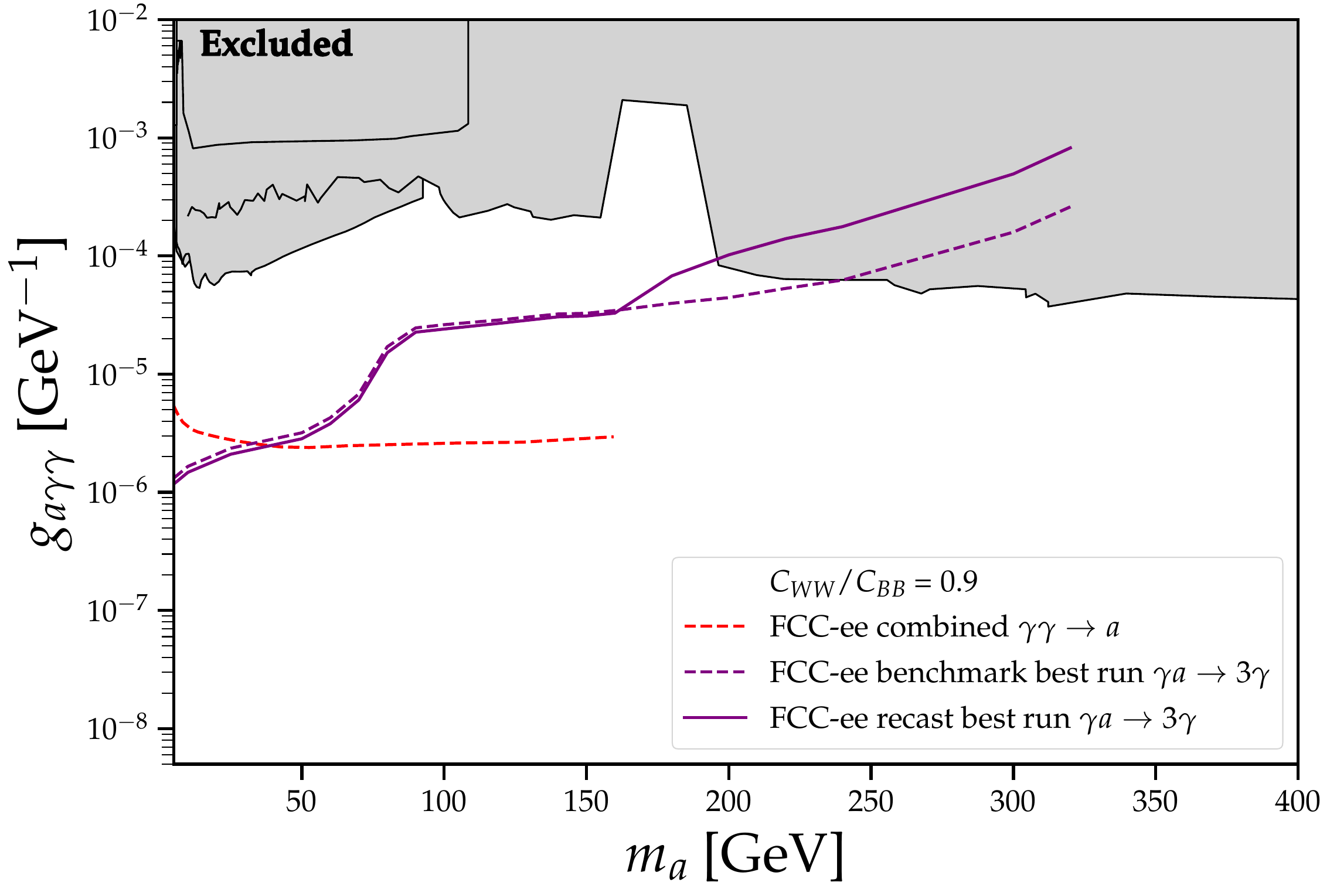}} 
    \subfigure[]{\includegraphics[width=0.45\textwidth]{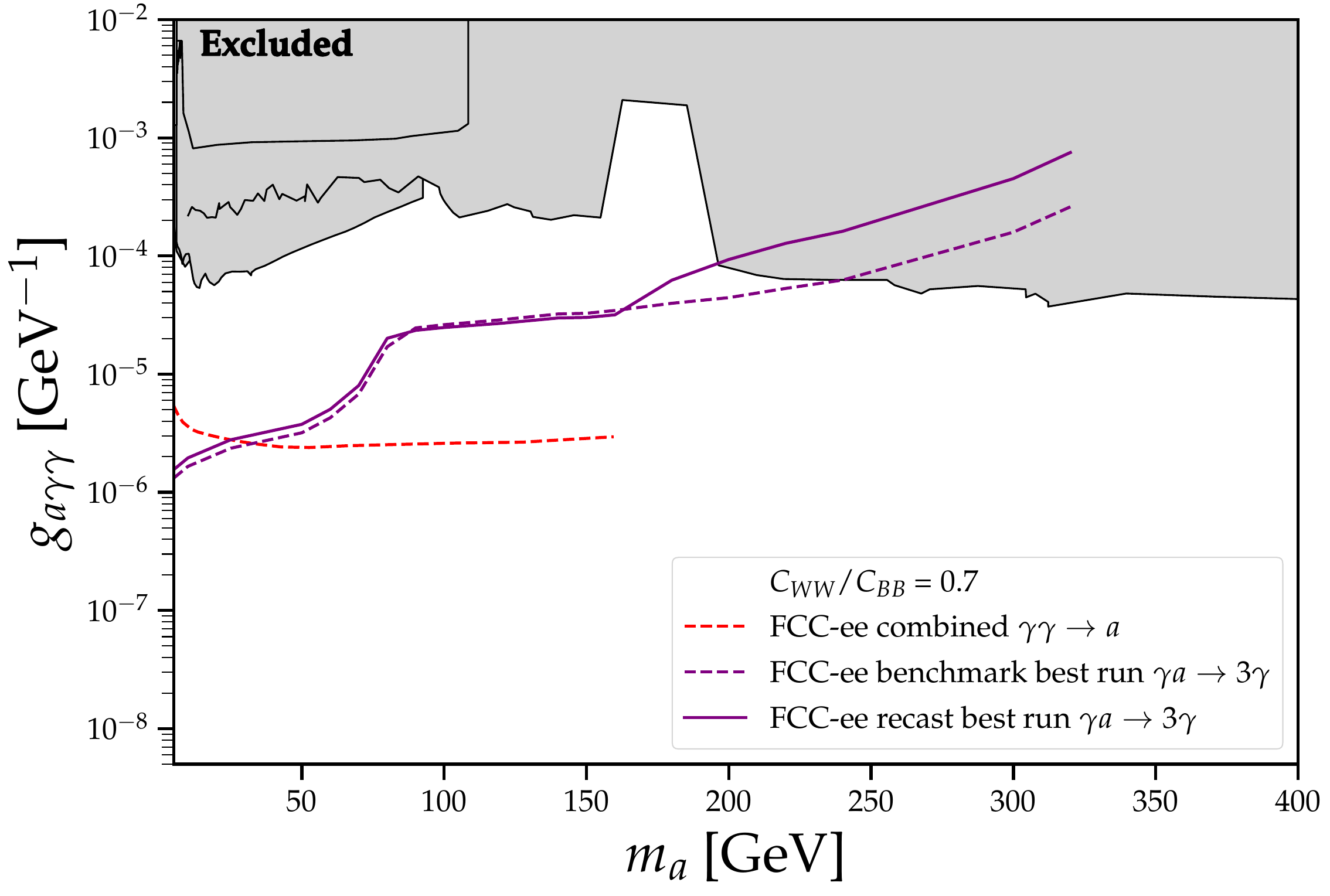}} 
    \subfigure[]{\includegraphics[width=0.45\textwidth]{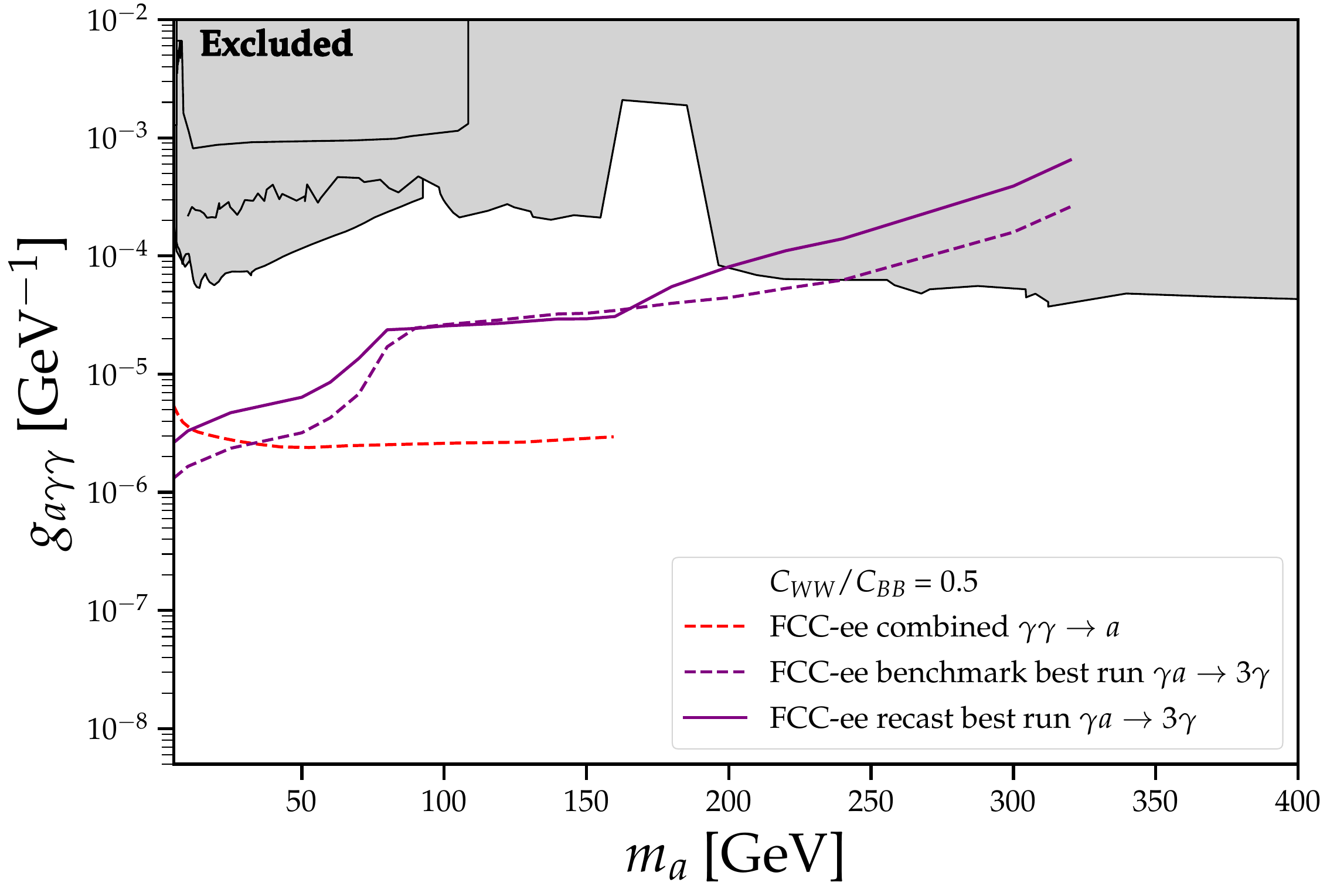}} 
    \subfigure[]{\includegraphics[width=0.45\textwidth]{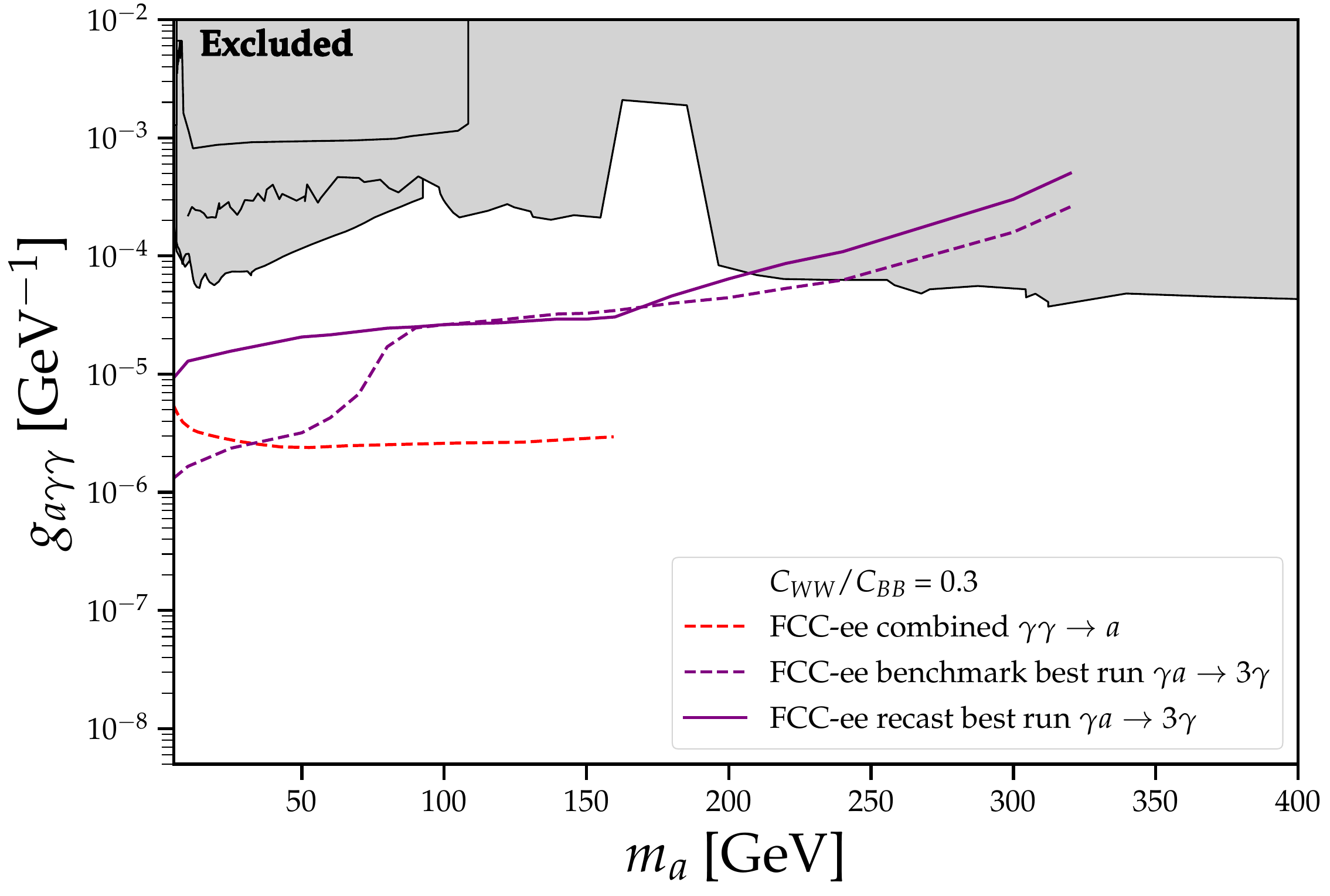}} 
    \subfigure[]{\includegraphics[width=0.45\textwidth]{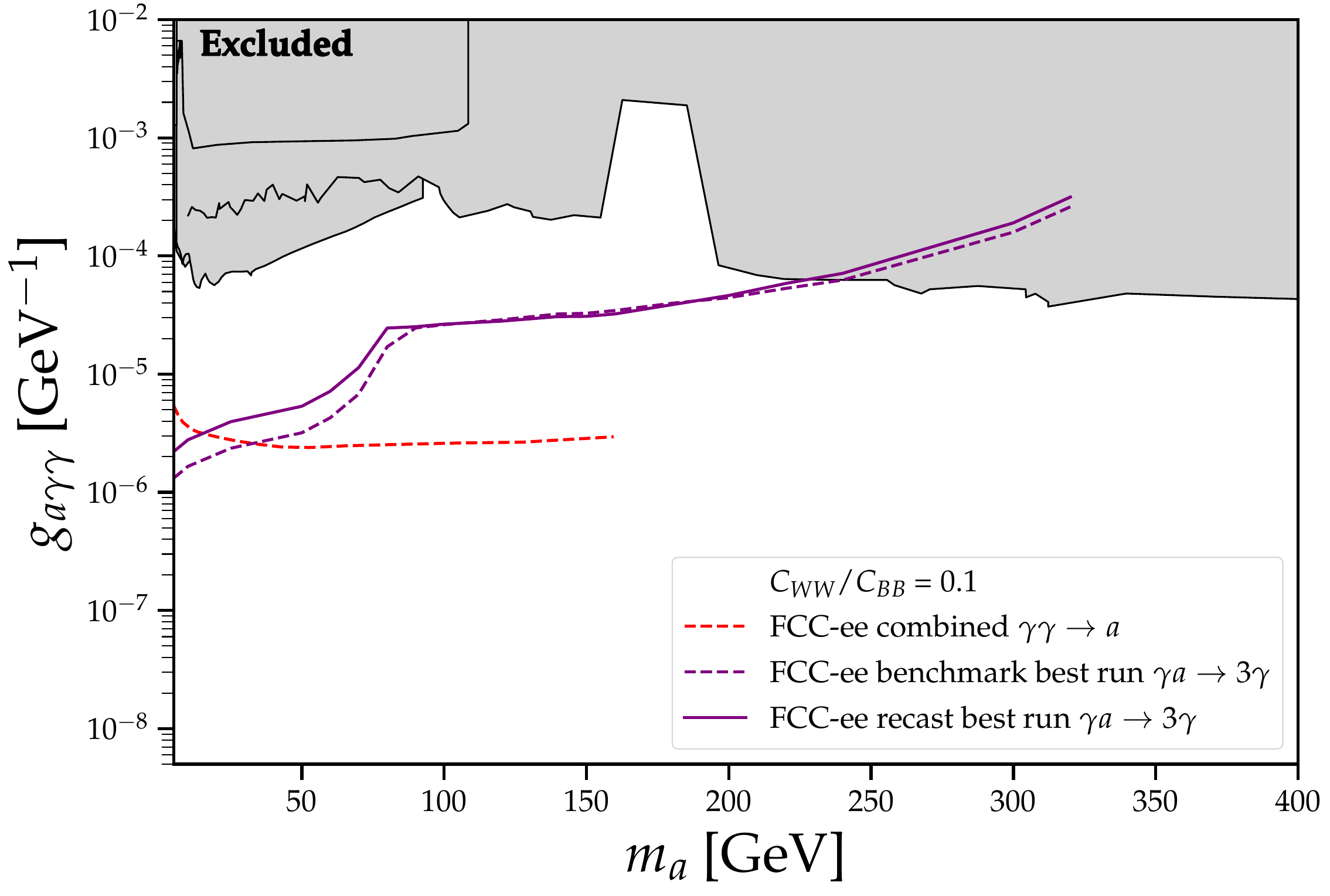}} 
    \subfigure[]{\includegraphics[width=0.45\textwidth]{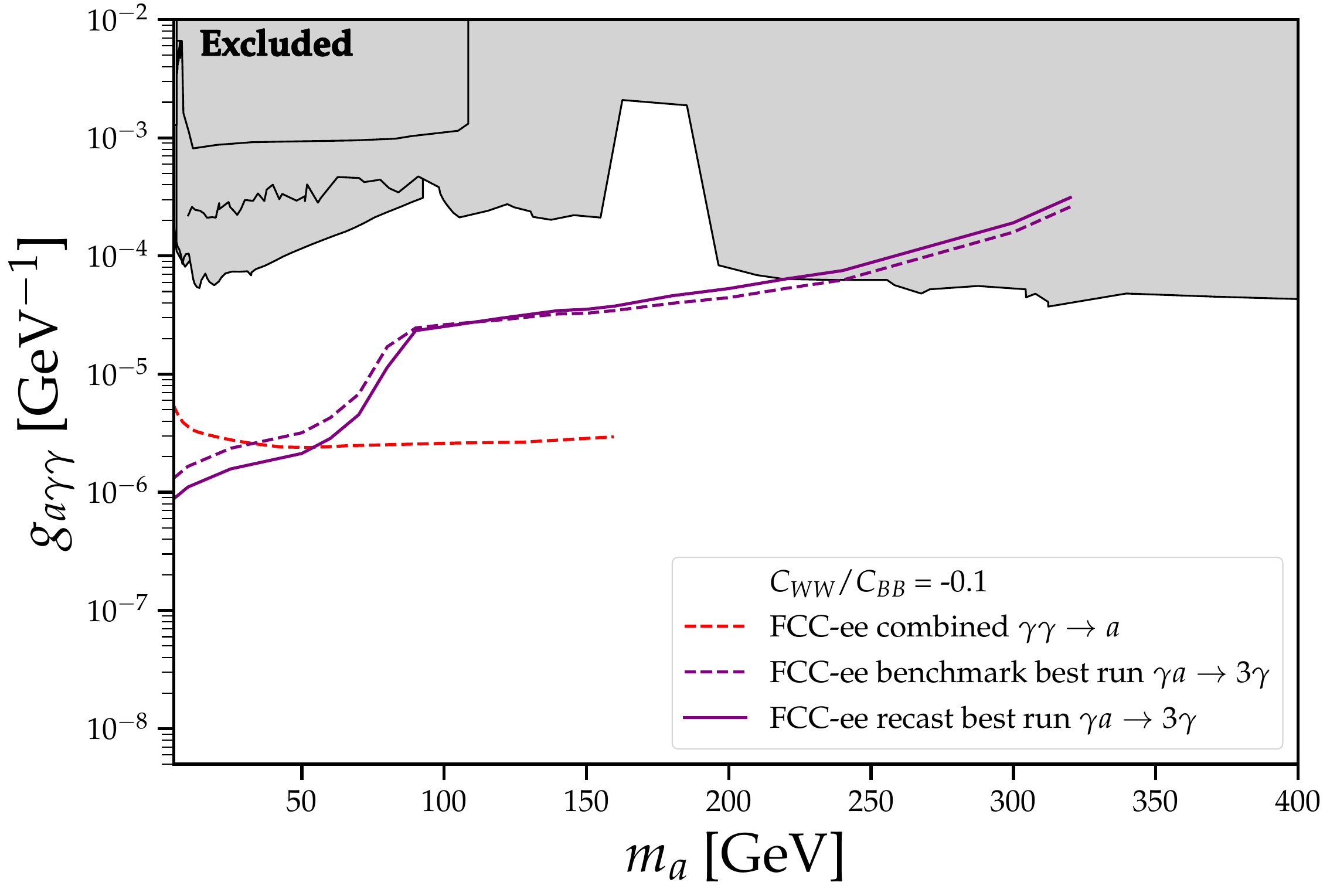}} 
    \caption{The 95\% CL projected sensitivity of the FCC-ee in the $g_{a\gamma\gamma}$-$m_a$ plane, assuming four $e^+ e^-$  interaction points. Plots a-f correspond to decreasing $C_{WW}/C_{BB}$ values from 0.9 to -0.1. The dashed purple line indicates the primary $3\gamma$ result, and the solid purple line corresponds to the recast $3\gamma$ result for the listed $C_{WW}/C_{BB}$ value. At each mass point, the strongest expected limit among the FCC-ee running stages is shown for the benchmark and recast $3\gamma$ results. The dashed red line indicates the $\gamma\gamma$-fusion~\cite{photon_fusion} result, for $m_a < m_Z$. The grey regions indicate the areas excluded by previous experiments~\cite{AxionLimits}, and are not altered.}
    \label{fig:recast_limits_positive_r}
\end{figure*}

\begin{figure*}[ht!]
    \centering
    \subfigure[]{\includegraphics[width=0.45\textwidth]{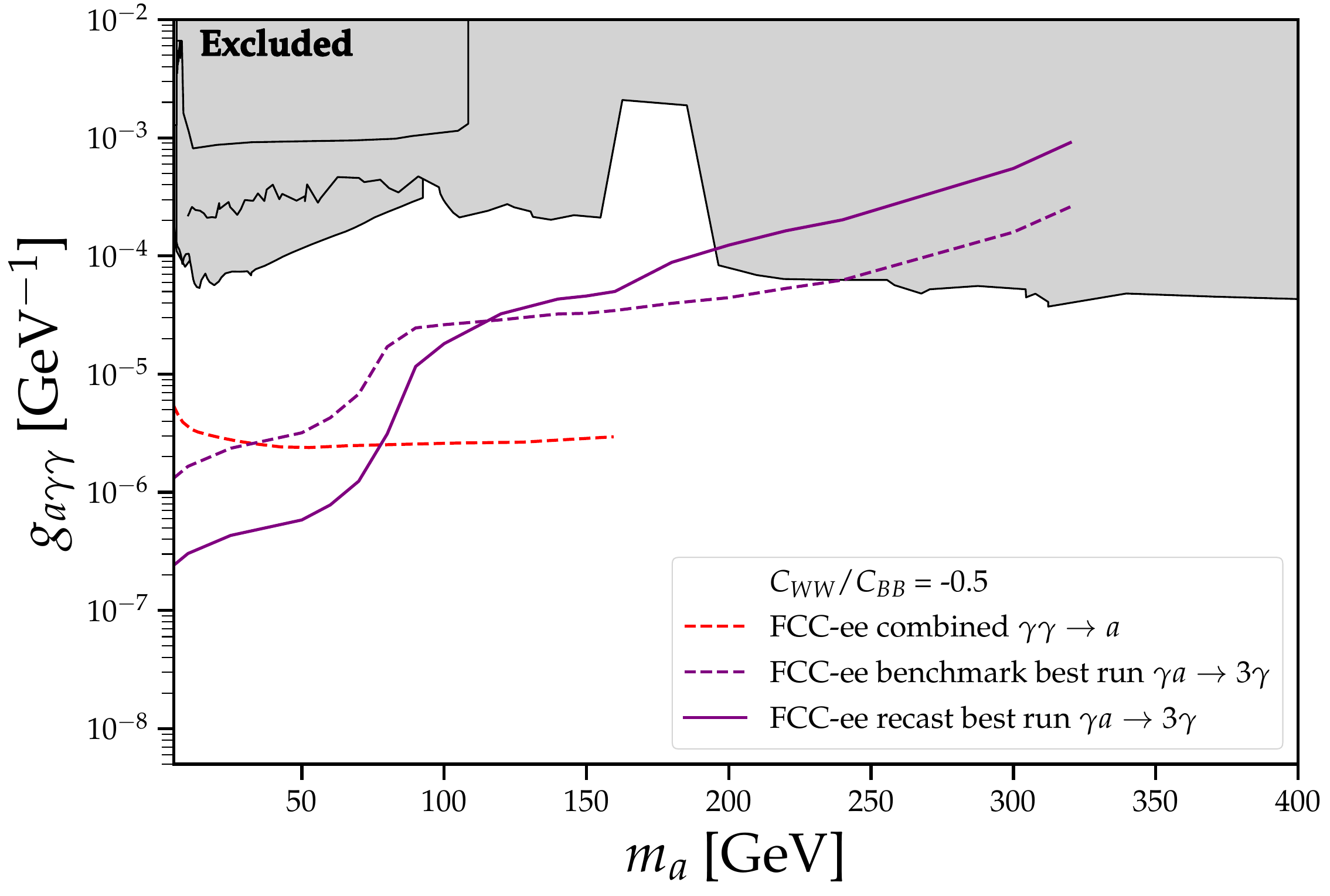}} 
    \subfigure[]{\includegraphics[width=0.45\textwidth]{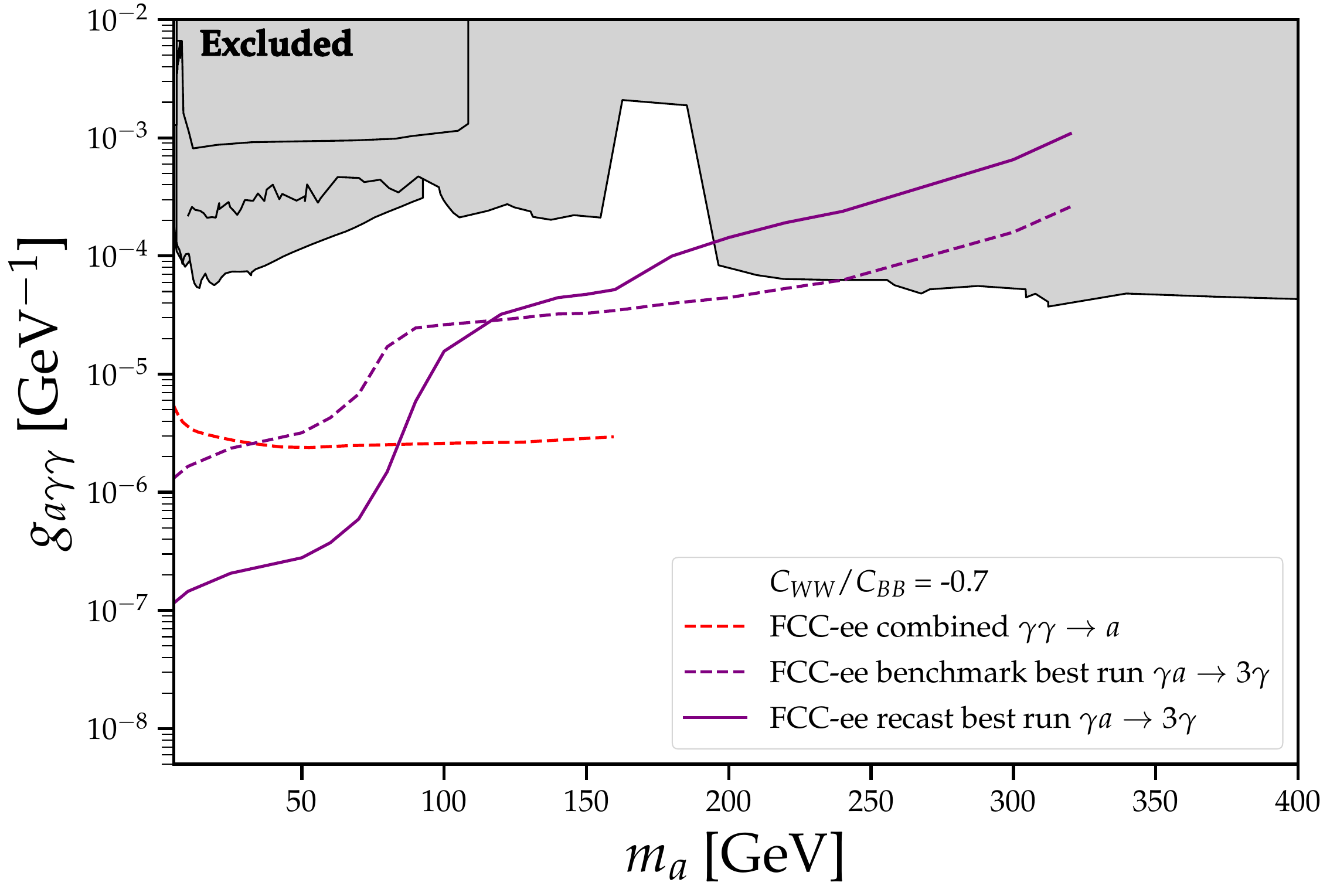}} 
    \subfigure[]{\includegraphics[width=0.45\textwidth]{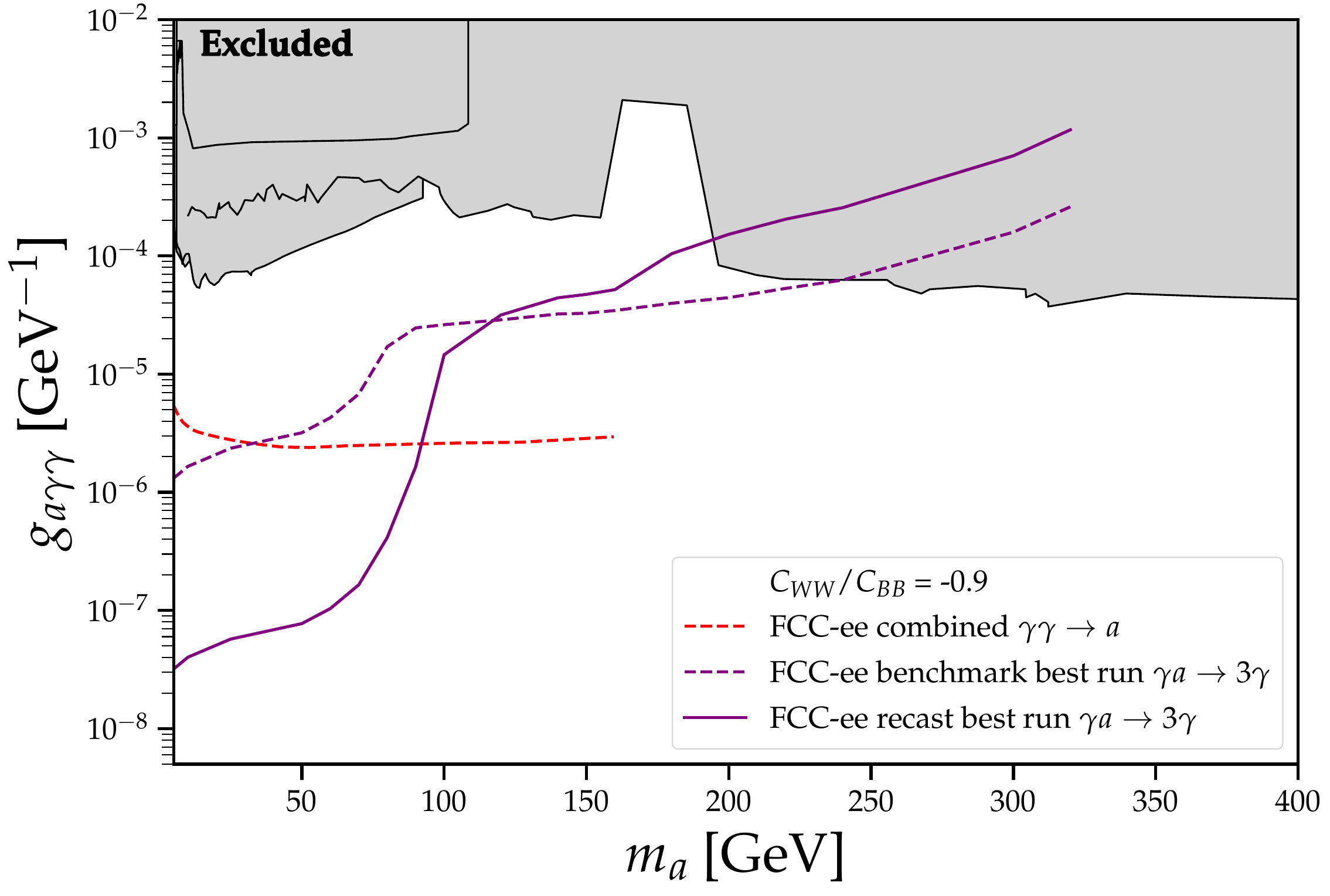}} 
    \subfigure[]{\includegraphics[width=0.45\textwidth]{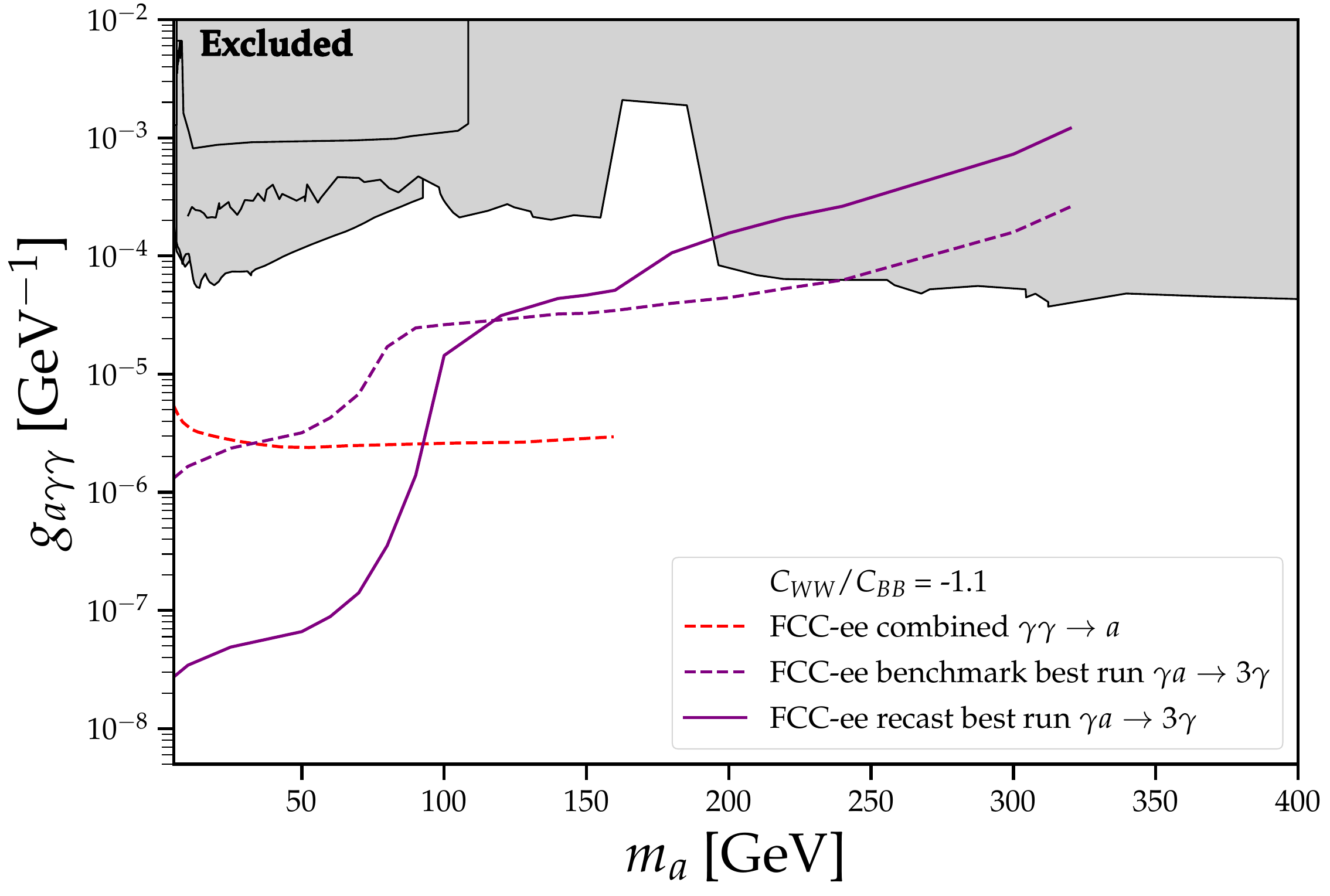}} 
    \subfigure[]{\includegraphics[width=0.45\textwidth]{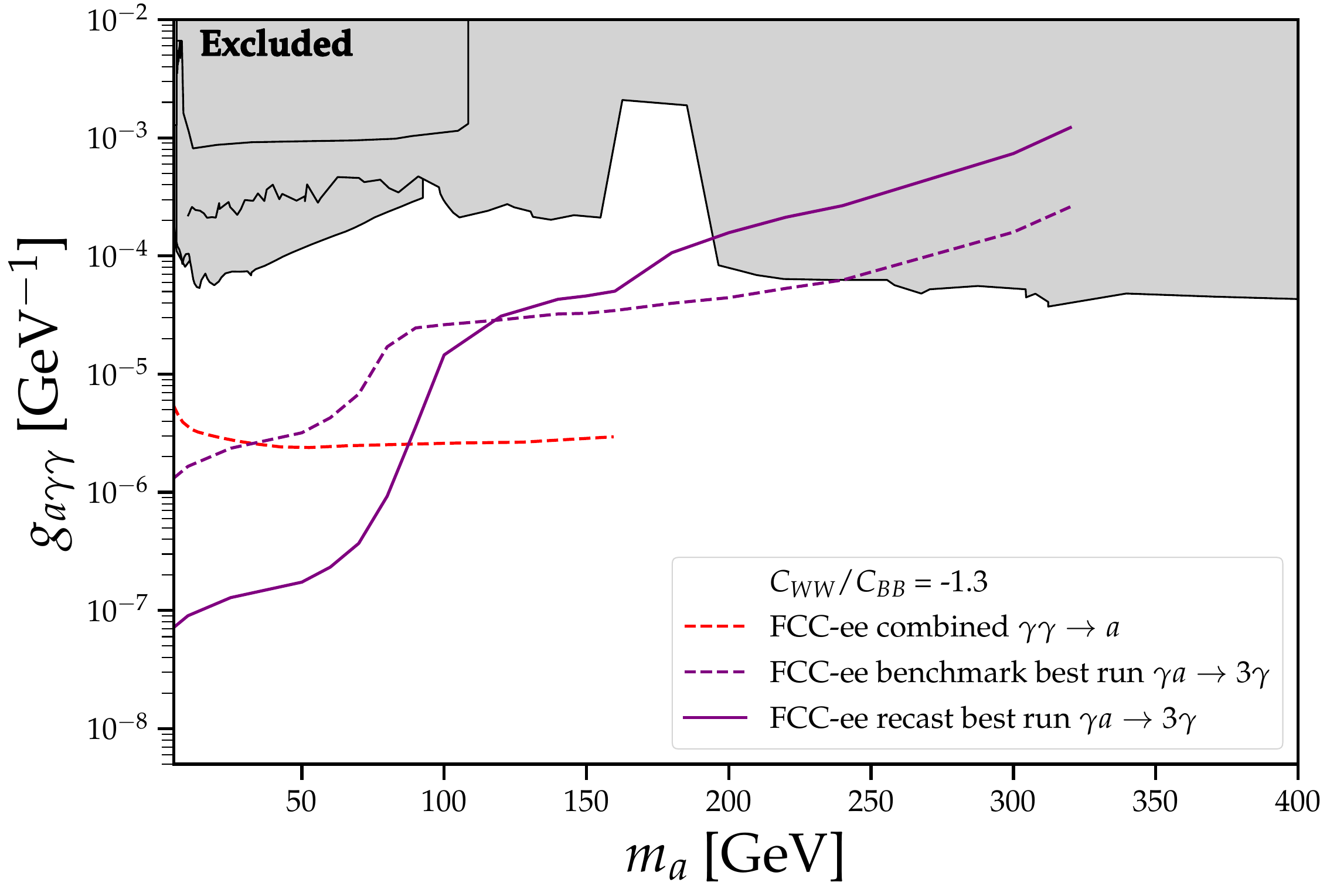}} 
    \subfigure[]{\includegraphics[width=0.45\textwidth]{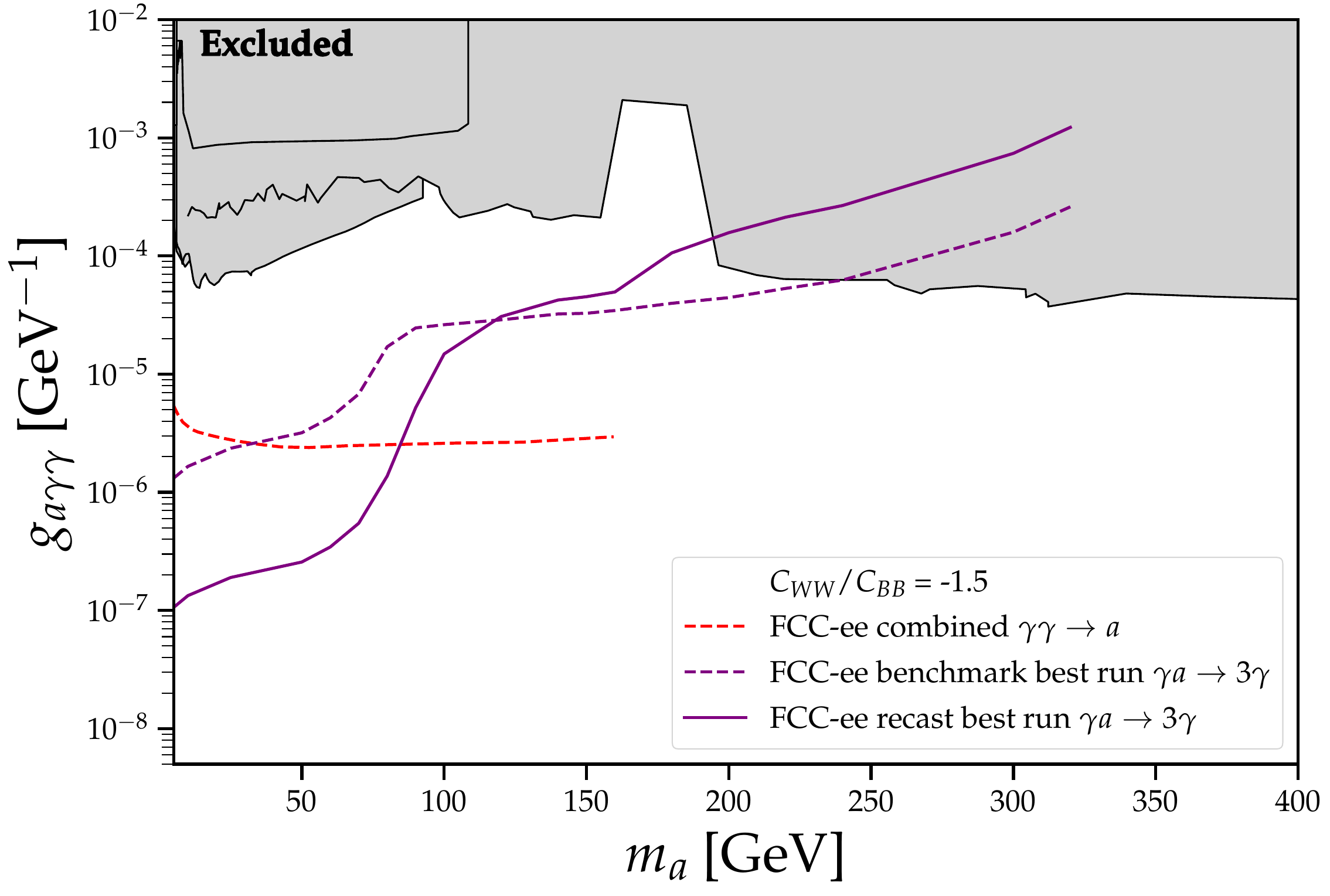}} 
    \caption{The 95\% CL projected sensitivity of the FCC-ee in the $g_{a\gamma\gamma}$-$m_a$ plane, assuming four $e^+ e^-$  interaction points. Plots a-f correspond to decreasing $C_{WW}/C_{BB}$ values from -0.5 to -1.5. The dashed purple line indicates the primary $3\gamma$ result, and the solid purple line corresponds to the recast $3\gamma$ result for the listed $C_{WW}/C_{BB}$ value. At each mass point, the strongest expected limit among the FCC-ee running stages is shown for the benchmark and recast $3\gamma$ results. The dashed red line indicates the $\gamma\gamma$-fusion~\cite{photon_fusion} result, for $m_a < m_Z$. The grey regions indicate the areas excluded by previous experiments~\cite{AxionLimits}, and are not altered.}
    \label{fig:recast_limits_negative_r}
\end{figure*}

The results for the $3\gamma$ final state are obtained for the benchmark scenario where $C_{BB} = 1$ and $C_{WW} = 0$, and they assume only tree-level couplings. For $m_a<m_Z$, this corresponds to a situation where $B(a\rightarrow\gamma\gamma)$ is 1. At the Z-pole running stage, the associated-production cross section is determined predominantly by the $C_{\gamma Z}$ coupling, which is related to $C_{\gamma\gamma}$ by the equation $C_{\gamma Z}=-s_w^2C_{\gamma\gamma}$. Away from the Z pole, the cross section depends on both $C_{\gamma\gamma}$ and $C_{\gamma Z}$. As $m_a$ increases, the additional decays into $\gamma Z$, $W^+W^-$, and $ZZ$ progressively become kinematically allowed. In the literature, for instance in Ref.~\cite{Cacciapaglia:2021agf}, different UV-complete models are proposed in which the choice of $C_{BB}$ and $C_{WW}$ differs from the benchmark adopted for this study. It is therefore useful to assess how the FCC-ee sensitivity changes with the relative size of $C_{BB}$ and $C_{WW}$, and to identify where associated production can provide information complementary to $\gamma\gamma$-fusion channels. The present analysis is recast for different choices of $r\equiv C_{WW}/C_{BB}$, varying between -1.5 and 1.5, and these results are shown in Figures~\ref{fig:recast_limits_positive_r} and~\ref{fig:recast_limits_negative_r}. For each value of $r$, the signal yield obtained in the benchmark scenario is rescaled using the associated-production cross section and B($a\to\gamma\gamma$). The total width used to determine this branching fraction includes the on-shell decays $a\to\gamma\gamma$, $a\to\gamma Z$, $a\to ZZ$, and, for nonzero $C_{WW}$, $a\to W^+W^-$, whenever kinematically allowed. Off-shell decay modes are not included. The selection efficiencies are kept unchanged, since for fixed $m_a$ and center-of-mass energy, the tree-level angular dependence of the associated-production process is unchanged.

\begin{figure*}[t]
    \centering
    {\includegraphics[width=0.75\textwidth]{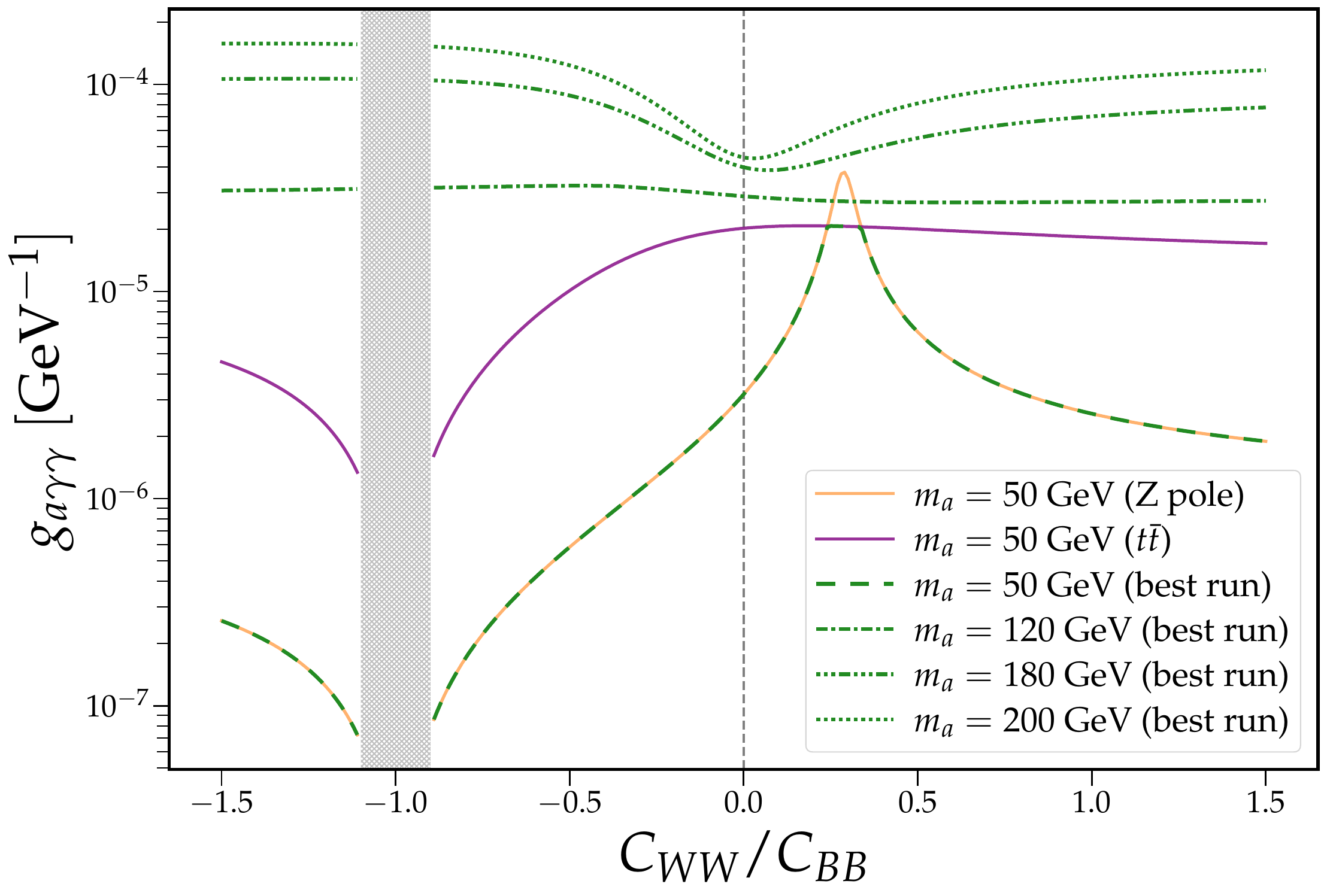}}
    \caption{The 95\% CL projected sensitivity of the FCC-ee, assuming four $e^+ e^-$  interaction points, as a function of $r\equiv C_{WW}/C_{BB}$. The yellow line corresponds to the $m_a = 50$ GeV results at the Z pole run, and the purple line corresponds to the $m_a = 50$ GeV results at the \ttbar threshold. The green lines show the best-run results for $m_a=50$, 120, 180, and 200 GeV. At each value of $r$, the strongest expected limit among the FCC-ee running stages is shown. The vertical dashed grey line indicates the primary result in this study. At the photophobic point $C_{WW} = -C_{BB}$, the tree-level $a\gamma\gamma$ coupling vanishes, and this recast has no sensitivity to the $3\gamma$ signal. The interval between $r=-1.1$ and -0.9 is shaded and not interpreted quantitatively, since radiative corrections become increasingly important near the photophobic point.}
    \label{fig:limits_vs_r}
\end{figure*}

The $3\gamma$ production is enhanced by the increase in $|C_{\gamma Z}|$ for $r<0$ and $r>0.9$, yielding a large gain in sensitivity for $m_{a}<m_{Z}$. The gain in sensitivity increases as one approaches the configuration $C_{WW}=-C_{BB}$, corresponding to the photophobic case described in Ref.~\cite{Cacciapaglia:2021agf}. Very close to this point, the tree-level $a\gamma\gamma$ coupling is strongly suppressed, and loop-induced contributions and lifetime effects can become model-dependent. The recast near the photophobic point should therefore be interpreted as an indication of the tree-level scaling rather than as a complete treatment of this parameter region. The $3\gamma$ production is decreased by the decrease in $C_{\gamma Z}$ for $r$ values between 0 and 0.9, with the largest dip in sensitivity corresponding to $r \approx 0.3$. For $m_{a}>m_{Z}$, the increase in $|C_{\gamma Z}|$ is compensated on one hand by the decrease in $C_{\gamma\gamma}$, and on the other hand by the fact that $B(a\rightarrow\gamma\gamma)$ becomes smaller than 1 because different decay channels open up. The resulting projected reach for $m_{a}>m_{Z}$ is therefore only weakly dependent on $r$. However, in this mass region, dedicated searches for decays into $\gamma Z$, $WW$, and $ZZ$ would provide additional measurements in case of discovery, thus constraining $C_{BB}$ and $C_{WW}$. To illustrate the dependence on $r$, Figure~\ref{fig:limits_vs_r} shows four benchmark masses: one below $m_Z$, a second between $m_Z$ and $2m_W$, a third between $2m_W$ and $2m_Z$, and a fourth above $2m_Z$. 
These choices separate the regions in which the on-shell $\gamma Z$, $W^+W^-$, and $ZZ$ decay modes successively become kinematically allowed. For $m_a < m_Z$, the results are sensitive to the value of $r$, and the variation in sensitivity spans many orders of magnitude. At $r = 0.3$, the drastic loss in sensitivity can be understood from Eq.~\ref{coefficient_relations}. At approximately this point, $C_{\gamma Z}$ is 0 and production of the $3\gamma$ final state occurs entirely via photon exchange, whose contribution is small at the Z pole run, but dominant at the other center-of-mass energies. This accounts for the difference in the variation in reach at the Z pole run compared to the \ttbar threshold results or the best-run results. However, for $m_a > m_Z$, the dependence of the projected reach on $r$ is substantially weaker than in the low-mass region.

Over the whole mass range of interest, two different production processes are available: associated production, which is studied in this paper, and $\gamma\gamma$ fusion. A complete study of their complementarity would require a recast of the results of Ref.~\cite{photon_fusion} in the same electroweak coupling basis used here. The results in Ref.~\cite{photon_fusion} are expressed in terms of an effective ALP coupling to photons and assume $B(a\to\gamma\gamma)=1$. For $m_{a}<m_{Z}$, this provides a useful approximation for comparison, since the additional on-shell electroweak decay modes are closed and the fusion sensitivity is governed primarily by the ALP-photon interaction. In this regime, the sensitivity in terms of $C_{\gamma\gamma}$, shown in Figures~\ref{fig:recast_limits_positive_r} and~\ref{fig:recast_limits_negative_r} as a red dashed line, is independent of the value of $r$. For $m_{a}>m_{Z}$, however, the same electroweak-gauge-boson Lagrangian also induces decays to $\gamma Z$, $ZZ$, and, for nonzero coupling to the weak gauge fields, $WW$. The branching fraction to two photons is then reduced by the opening of the additional decay modes, while additional electroweak fusion processes can also contribute to production. A consistent comparison in this high-mass region would therefore require a dedicated recalculation of the fusion-channel sensitivity, as well as a corresponding recast of existing limits. This is beyond the scope of the present study.

\section{Summary and outlook}

We have presented a feasibility study of the sensitivity to axion-like particles (ALPs) at the FCC-ee, a planned future $e^+ e^-$ collider. The ALP is produced in association with a photon and decays to two photons, $e^+ e^- \to a \gamma \to 3\gamma$. The process is simulated using the IDEA detector configuration at all proposed FCC-ee center-of-mass energies: the Z pole, WW, ZH, and \ttbar running stages. Event selection criteria are applied to optimize the significance, and sensitivity projections are made for all runs separately, as well as all runs combined. As expected, the Z pole run probes the smallest couplings due to the expected high luminosity. In contrast, the higher center-of-mass energy runs can probe larger ALP masses, specifically from around 90 to 320 GeV. The sensitivity of all runs combined is improved with respect to the individual FCC-ee runs for masses between 80 and 150 GeV. For masses between 90 and 300 GeV, the $3\gamma$ results improve upon the current LHC limits. The overall result is mass-dependent because the photon assignment and optimized final selection are evaluated separately for each tested ALP mass hypothesis. The results include only the irreducible three-photon background and do not include systematic uncertainties, reducible backgrounds, or detector-specific beam-related effects. Overall, this study shows that there is good potential to explore the ALP phase space at the FCC-ee by extending to higher center-of-mass energy runs.

The associated-production and photon-fusion channels depend on different combinations of the ALP couplings to the electroweak gauge fields, so measurements of both channels would help disentangle the underlying electroweak structure of the ALP interactions. Associated production in the $3\gamma$ final state is sensitive to the value of $r \equiv C_{WW}/C_{BB}$ for $m_a < m_Z$, with the largest gain in sensitivity as the photophobic case $C_{WW} = -C_{BB}$ is approached. For $m_a > m_Z$, the result is weakly dependent on the value of $r$. For $m_a < m_Z$, comparisons have been made to previous ALP studies at the FCC-ee. In particular, the $\gamma\gamma$-fusion result is governed primarily by $C_{\gamma\gamma}$, while the $3\gamma$ final state is also sensitive to $C_{\gamma Z}$, providing complementary information on $C_{WW}/C_{BB}$. In the higher mass region, additional electroweak fusion processes can contribute to production, and a consistent combination of these channels would provide further constraints on $C_{WW}$ and $C_{BB}$. In summary, we demonstrate the strong potential of the FCC-ee to probe ALPs and to help constrain the underlying electroweak structure of their couplings.

\bmhead{Acknowledgements} JA, EB, FB, JDA, CD, JN, AP, and LR acknowledge support from DESY (Hamburg, Germany), a member of the Helmholtz Association HGF, and support by the Deutsche Forschungsgemeinschaft (DFG, German Research Foundation) under Germany’s Excellence Strategy -- EXC 2121 ``Quantum Universe'' -- 390833306. JDA acknowledges partial support from the Egyptian governmental funding agency, Science, Technology \& Innovation Funding Authority (STDF) (Grant No. 50806).

\bmhead{Statements and Declarations} The authors have no competing interests to declare that are relevant to the content of this article.

\begin{appendices}

\section{}

This appendix contains tables with detailed information regarding the event normalization and selection. There is a table that contains the cross sections used to normalize the number of signal events. Another table contains the values for signal efficiency and background rejection for each selection stage. The rest of the tables contain detailed information on the event selection at each center-of-mass energy, as well as the number of events after the preselection and final selection, for all FCC-ee runs.

\begin{table*}[tpb]
\begin{center}
\begin{tabular}{ c || c | c | c | c }
 & \multicolumn{4}{c}{$\sigma(e^+ e^- \to \gamma a)$ [pb]} \\
 \hline\hline
 $m_a$ [GeV] & Z pole & WW & ZH & \ttbar \\
 \hline\hline
 5   & $2.4 \times 10^0$ & $2.2 \times 10^{-2}$ & $2.0 \times 10^{-2}$ & $2.0 \times 10^{-2}$ \\
 \hline
 10  & $2.4 \times 10^0$ & $2.2 \times 10^{-2}$ & $2.0 \times 10^{-2}$ & $2.0 \times 10^{-2}$ \\
 \hline
 25  & $2.0 \times 10^0$ & $2.0 \times 10^{-2}$ & $2.0 \times 10^{-2}$ & $2.0 \times 10^{-2}$ \\
 \hline
 50  & $8.5 \times 10^{-1}$ & $1.6 \times 10^{-2}$ & $1.8 \times 10^{-2}$ & $1.9 \times 10^{-2}$ \\
 \hline
 60  & $4.5 \times 10^{-1}$ & $1.4 \times 10^{-2}$ & $1.7 \times 10^{-2}$ & $1.9 \times 10^{-2}$ \\
 \hline
 70  & $1.7 \times 10^{-1}$ & $1.2 \times 10^{-2}$ & $1.6 \times 10^{-2}$ & $1.8 \times 10^{-2}$ \\
 \hline
 80  & $3.0 \times 10^{-2}$ & $9.2 \times 10^{-3}$ & $1.4 \times 10^{-2}$ & $1.7 \times 10^{-2}$ \\
 \hline
 90  & ---   & $7.0 \times 10^{-3}$ & $1.3 \times 10^{-2}$ & $1.7 \times 10^{-2}$ \\
 \hline
 100 & ---   & $4.9 \times 10^{-3}$ & $1.2 \times 10^{-2}$ & $1.6 \times 10^{-2}$ \\
 \hline
 120 & ---   & $1.7 \times 10^{-3}$ & $8.3 \times 10^{-3}$ & $1.4 \times 10^{-2}$ \\
 \hline
 140 & ---   & $2.5 \times 10^{-4}$ & $5.3 \times 10^{-3}$ & $1.1 \times 10^{-2}$ \\
 \hline
 150 & ---   & $3.4 \times 10^{-5}$ & $4.1 \times 10^{-3}$ & $1.0 \times 10^{-2}$ \\
 \hline
 160 & ---   & ---   & $3.0 \times 10^{-3}$ & $9.1 \times 10^{-3}$ \\
 \hline
 180 & ---   & ---   & $1.4 \times 10^{-3}$ & $7.1 \times 10^{-3}$ \\
 \hline
 200 & ---   & ---   & $4.5 \times 10^{-4}$ & $5.4 \times 10^{-3}$ \\
 \hline
 220 & ---   & ---   & $6.2 \times 10^{-5}$ & $3.9 \times 10^{-3}$ \\
 \hline
 240 & ---   & ---   & ---   & $2.7 \times 10^{-3}$ \\
 \hline
 300 & ---   & ---   & ---   & $4.7 \times 10^{-4}$ \\
 \hline
 320 & ---   & ---   & ---   & $1.7 \times 10^{-4}$ \\

\end{tabular}
\caption{The ALP production cross section for each mass point and center-of-mass energy. The cross sections are computed with \texttt{MG5aMC@NLO} v. 3.6.3.}
\label{tab:cross_sections}
\end{center}
\end{table*}

\begin{table*}[tpb]
\begin{center}
\begin{tabular}{ c | c || c | c | c | c || c | c | c | c }
 $m_a$ & & \multicolumn{4}{c||}{After preselection} & \multicolumn{4}{c}{After final selection} \\ \hhline{~~--------}
  {[GeV]} & & Z pole & WW & ZH & \ttbar & Z pole & WW & ZH & \ttbar \\ 
 \hline 
 \multirow{2}{*}{5} & S efficiency & 0.826 & 0.886 & 0.902 & 0.914 & 0.517 & 0.437 & 0.465 & 0.473 \\
  & B rejection & 0.536 & 0.484 & 0.453 & 0.423 & 0.997 & 0.999 & 0.999 & 0.999 \\ 
 \hline
 \multirow{2}{*}{10} & S efficiency & 0.829 & 0.879 & 0.894 & 0.904 & 0.494 & 0.462 & 0.462 & 0.548 \\
  & B rejection & 0.536 & 0.484 & 0.453 & 0.423 & 0.996 & 0.998 & 0.998 & 0.998 \\ 
 \hline
 \multirow{2}{*}{25} & S efficiency & 0.875 & 0.902 & 0.893 & 0.895 & 0.542 & 0.493 & 0.508 & 0.469 \\
  & B rejection & 0.536 & 0.483 & 0.453 & 0.423 & 0.996 & 0.997 & 0.995 & 0.998 \\ 
 \hline
 \multirow{2}{*}{50} & S efficiency & 0.874 & 0.900 & 0.900 & 0.901 & 0.387 & 0.520 & 0.589 & 0.523 \\
  & B rejection & 0.536 & 0.483 & 0.453 & 0.423 & 0.999 & 0.997 & 0.996 & 0.997 \\ 
 \hline
 \multirow{2}{*}{60} & S efficiency & 0.874 & 0.901 & 0.900 & 0.900 & 0.417 & 0.474 & 0.611 & 0.518 \\
  & B rejection & 0.536 & 0.483 & 0.452 & 0.423 & 0.999 & 0.998 & 0.996 & 0.997 \\ 
 \hline
 \multirow{2}{*}{70} & S efficiency & 0.875 & 0.901 & 0.900 & 0.900 & 0.404 & 0.500 & 0.640 & 0.555 \\
  & B rejection & 0.536 & 0.483 & 0.452 & 0.423 & 0.999 & 0.998 & 0.996 & 0.996 \\ 
 \hline
 \multirow{2}{*}{80} & S efficiency & 0.878 & 0.901 & 0.900 & 0.900 & 0.406 & 0.410 & 0.561 & 0.585 \\
  & B rejection & 0.536 & 0.483 & 0.452 & 0.423 & 0.999 & 0.999 & 0.997 & 0.996 \\ 
 \hline
 \multirow{2}{*}{90} & S efficiency & \multirow{2}{*}{---} & 0.901 & 0.900 & 0.900 & \multirow{2}{*}{---} & 0.434 & 0.606 & 0.550 \\
  & B rejection &  & 0.483 & 0.452 & 0.423 &  & 0.999 & 0.997 & 0.997 \\ 
 \hline
 \multirow{2}{*}{100} & S efficiency & \multirow{2}{*}{---} & 0.901 & 0.900 & 0.900 & \multirow{2}{*}{---} & 0.432 & 0.601 & 0.592 \\
  & B rejection &  & 0.483 & 0.452 & 0.423 &  & 0.999 & 0.997 & 0.997 \\ 
 \hline
 \multirow{2}{*}{120} & S efficiency & \multirow{2}{*}{---} & 0.901 & 0.901 & 0.901 & \multirow{2}{*}{---} & 0.474 & 0.606 & 0.604 \\
  & B rejection &  & 0.483 & 0.452 & 0.423 &  & 0.999 & 0.998 & 0.997 \\ 
 \hline
 \multirow{2}{*}{140} & S efficiency & \multirow{2}{*}{---} & 0.901 & 0.901 & 0.901 & \multirow{2}{*}{---} & 0.519 & 0.627 & 0.620 \\
  & B rejection &  & 0.483 & 0.452 & 0.423 &  & 0.998 & 0.997 & 0.997 \\ 
 \hline
 \multirow{2}{*}{150} & S efficiency & \multirow{2}{*}{---} & 0.905 & 0.901 & 0.900 & \multirow{2}{*}{---} & 0.456 & 0.607 & 0.520 \\
  & B rejection &  & 0.483 & 0.452 & 0.423 &  & 0.998 & 0.998 & 0.998 \\ 
 \hline
 \multirow{2}{*}{160} & S efficiency & \multirow{2}{*}{---} & \multirow{2}{*}{---} & 0.902 & 0.901 & \multirow{2}{*}{---} & \multirow{2}{*}{---} & 0.662 & 0.484 \\
  & B rejection &  &  & 0.452 & 0.423 &  &  & 0.998 & 0.999 \\ 
 \hline
 \multirow{2}{*}{180} & S efficiency & \multirow{2}{*}{---} & \multirow{2}{*}{---} & 0.902 & 0.901 & \multirow{2}{*}{---} & \multirow{2}{*}{---} & 0.714 & 0.507 \\
  & B rejection &  &  & 0.452 & 0.423 &  &  & 0.997 & 0.999 \\ 
 \hline
 \multirow{2}{*}{200} & S efficiency & \multirow{2}{*}{---} & \multirow{2}{*}{---} & 0.901 & 0.901 & \multirow{2}{*}{---} & \multirow{2}{*}{---} & 0.683 & 0.433 \\
  & B rejection &  &  & 0.452 & 0.423 &  &  & 0.998 & 0.999 \\ 
 \hline
 \multirow{2}{*}{220} & S efficiency & \multirow{2}{*}{---} & \multirow{2}{*}{---} & 0.904 & 0.901 & \multirow{2}{*}{---} & \multirow{2}{*}{---} & 0.685 & 0.514 \\
  & B rejection &  &  & 0.452 & 0.423 &  &  & 0.997 & 0.999 \\ 
 \hline
 \multirow{2}{*}{240} & S efficiency & \multirow{2}{*}{---} & \multirow{2}{*}{---} & \multirow{2}{*}{---} & 0.901 & \multirow{2}{*}{---} & \multirow{2}{*}{---} & \multirow{2}{*}{---} & 0.452 \\
  & B rejection &  &  &  & 0.423 &  &  &  & 0.999 \\ 
 \hline
 \multirow{2}{*}{300} & S efficiency & \multirow{2}{*}{---} & \multirow{2}{*}{---} & \multirow{2}{*}{---} & 0.902 & \multirow{2}{*}{---} & \multirow{2}{*}{---} & \multirow{2}{*}{---} & 0.529 \\
  & B rejection &  &  &  & 0.423 &  &  &  & 0.999 \\ 
 \hline
 \multirow{2}{*}{320} & S efficiency & \multirow{2}{*}{---} & \multirow{2}{*}{---} & \multirow{2}{*}{---} & 0.902 & \multirow{2}{*}{---} & \multirow{2}{*}{---} & \multirow{2}{*}{---} & 0.512 \\
  & B rejection &  &  &  & 0.423 &  &  &  & 0.999 \\ 
\end{tabular}
\caption{The signal (S) efficiency and background (B) rejection are shown for each mass point. The first two columns correspond to $m_a$ and S efficiency/B rejection. The next four columns correspond to the various center-of-mass energies studied, with values after the preselection is applied. The last four columns correspond to the various center-of-mass energies once the final selection is applied.}
\label{seleff_backrej}
\end{center}
\end{table*}

\begin{table*}[tbp]
\begin{center}
\resizebox{\linewidth}{!}{
\begin{tabular}{ c || c | c | c | c || c | c | c | c }
\multicolumn{9}{c}{Z pole running stage} \\
\hline\hline
 $m_a$ & \multirow{2}{*}{$M_{cut}$} & \multirow{2}{*}{$|\cos{\theta_{\gamma_1}}|$} & \multirow{2}{*}{$|\phi_{\gamma_1}|$} & \multirow{2}{*}{$\Delta \alpha_{\gamma_1 \gamma_2}$} & \multirow{2}{*}{} & \multirow{2}{*}{No selection} & \multirow{2}{*}{Preselection} & \multirow{2}{*}{Final selection} \\ [0.0ex]
 [GeV] & & & & & & & & \\ 
 \hline 
 \multirow{2}{*}{5} & \multirow{2}{*}{\textless 2.1} & \multirow{2}{*}{\textless 0.7} & \multirow{2}{*}{\textless 1.6} & \multirow{2}{*}{\textless 0.3} & S & $2.4 \times 10^{8}$ $\pm$ $2.4 \times 10^{5}$ & $2.0 \times 10^{8}$ $\pm$ $2.2 \times 10^{5}$ & $1.3 \times 10^{8}$ $\pm$ $1.8 \times 10^{5}$ \\ 
  & & & & & B & $1.8 \times 10^{9}$ $\pm$ $1.8 \times 10^{6}$ & $8.6 \times 10^{8}$ $\pm$ $1.3 \times 10^{6}$ & $5.4 \times 10^{6}$ $\pm$ $1.0 \times 10^{5}$ \\ 
 \hline
 \multirow{2}{*}{10} & \multirow{2}{*}{\textless 1.9} & \multirow{2}{*}{\textless 0.7} & \multirow{2}{*}{\textless 1.6} & \multirow{2}{*}{\textless 0.6} & S & $2.4 \times 10^{8}$ $\pm$ $2.4 \times 10^{5}$ & $2.0 \times 10^{8}$ $\pm$ $2.2 \times 10^{5}$ & $1.2 \times 10^{8}$ $\pm$ $1.7 \times 10^{5}$ \\ 
  & & & & & B & $1.8 \times 10^{9}$ $\pm$ $1.8 \times 10^{6}$ & $8.6 \times 10^{8}$ $\pm$ $1.3 \times 10^{6}$ & $7.2 \times 10^{6}$ $\pm$ $1.2 \times 10^{5}$ \\ 
 \hline
 \multirow{2}{*}{25} & \multirow{2}{*}{\textless 1.8} & \multirow{2}{*}{\textless 0.8} & \multirow{2}{*}{\textless 1.6} & \multirow{2}{*}{\textless 1.5} & S & $1.9 \times 10^{8}$ $\pm$ $1.9 \times 10^{5}$ & $1.7 \times 10^{8}$ $\pm$ $1.8 \times 10^{5}$ & $1.0 \times 10^{8}$ $\pm$ $1.4 \times 10^{5}$ \\ 
  & & & & & B & $1.8 \times 10^{9}$ $\pm$ $1.8 \times 10^{6}$ & $8.6 \times 10^{8}$ $\pm$ $1.3 \times 10^{6}$ & $7.4 \times 10^{6}$ $\pm$ $1.2 \times 10^{5}$ \\ 
 \hline
 \multirow{2}{*}{50} & \multirow{2}{*}{\textless 2.0} & \multirow{2}{*}{\textless 0.8} & \multirow{2}{*}{\textless 1.1} & \multirow{2}{*}{\textless 2.3} & S & $8.4 \times 10^{7}$ $\pm$ $8.4 \times 10^{4}$ & $7.3 \times 10^{7}$ $\pm$ $7.9 \times 10^{4}$ & $3.2 \times 10^{7}$ $\pm$ $5.2 \times 10^{4}$ \\ 
  & & & & & B & $1.8 \times 10^{9}$ $\pm$ $1.8 \times 10^{6}$ & $8.6 \times 10^{8}$ $\pm$ $1.3 \times 10^{6}$ & $2.0 \times 10^{6}$ $\pm$ $6.1 \times 10^{4}$ \\ 
 \hline
 \multirow{2}{*}{60} & \multirow{2}{*}{\textless 2.0} & \multirow{2}{*}{\textless 0.7} & \multirow{2}{*}{\textless 1.2} & \multirow{2}{*}{\textless 2.6} & S & $4.5 \times 10^{7}$ $\pm$ $4.5 \times 10^{4}$ & $3.9 \times 10^{7}$ $\pm$ $4.2 \times 10^{4}$ & $1.9 \times 10^{7}$ $\pm$ $2.9 \times 10^{4}$ \\ 
  & & & & & B & $1.8 \times 10^{9}$ $\pm$ $1.8 \times 10^{6}$ & $8.6 \times 10^{8}$ $\pm$ $1.3 \times 10^{6}$ & $2.1 \times 10^{6}$ $\pm$ $6.3 \times 10^{4}$ \\ 
 \hline
 \multirow{2}{*}{70} & \multirow{2}{*}{\textless 1.8} & \multirow{2}{*}{\textless 0.8} & \multirow{2}{*}{\textless 1.1} & \multirow{2}{*}{\textless 2.9} & S & $1.7 \times 10^{7}$ $\pm$ $1.7 \times 10^{4}$ & $1.5 \times 10^{7}$ $\pm$ $1.6 \times 10^{4}$ & $6.9 \times 10^{6}$ $\pm$ $1.1 \times 10^{4}$ \\ 
  & & & & & B & $1.8 \times 10^{9}$ $\pm$ $1.8 \times 10^{6}$ & $8.6 \times 10^{8}$ $\pm$ $1.3 \times 10^{6}$ & $2.0 \times 10^{6}$ $\pm$ $6.1 \times 10^{4}$ \\ 
 \hline
 \multirow{2}{*}{80} & \multirow{2}{*}{\textless 1.8} & \multirow{2}{*}{\textless 0.8} & \multirow{2}{*}{\textless 1.1} & \multirow{2}{*}{\textless 3.1} & S & $3.0 \times 10^{6}$ $\pm$ $3.0 \times 10^{3}$ & $2.6 \times 10^{6}$ $\pm$ $2.8 \times 10^{3}$ & $1.2 \times 10^{6}$ $\pm$ $1.9 \times 10^{3}$ \\ 
  & & & & & B & $1.8 \times 10^{9}$ $\pm$ $1.8 \times 10^{6}$ & $8.6 \times 10^{8}$ $\pm$ $1.3 \times 10^{6}$ & $2.6 \times 10^{6}$ $\pm$ $7.0 \times 10^{4}$ \\ 
\end{tabular}
}
\caption{Event selections at the Z pole running stage, for each mass point. The rightmost columns show the number of events for no applied selection, after the preselection criteria, and after the final selection criteria, for both signal (S) and background (B).}
\label{zpole_selections}
\end{center}
\end{table*}

\begin{table*}[tbp]
\begin{center}
\resizebox{\linewidth}{!}{
\begin{tabular}{ c || c | c | c | c || c | c | c | c }
\multicolumn{9}{c}{WW running stage} \\
\hline\hline
 $m_a$ & \multirow{2}{*}{$M_{cut}$} & \multirow{2}{*}{$|\cos{\theta_{\gamma_1}}|$} & \multirow{2}{*}{$|\phi_{\gamma_1}|$} & \multirow{2}{*}{$\Delta \alpha_{\gamma_1 \gamma_2}$} & \multirow{2}{*}{} & \multirow{2}{*}{No selection} & \multirow{2}{*}{Preselection} & \multirow{2}{*}{Final selection} \\ [0.0ex]
 [GeV] & & & & & & & & \\ 
 \hline 
 \multirow{2}{*}{5} & \multirow{2}{*}{\textless 2.1} & \multirow{2}{*}{\textless 0.6} & \multirow{2}{*}{\textless 1.6} & \multirow{2}{*}{\textless 0.2} & S & $4.1 \times 10^{5}$ $\pm$ $4.1 \times 10^{2}$ & $3.7 \times 10^{5}$ $\pm$ $3.9 \times 10^{2}$ & $1.8 \times 10^{5}$ $\pm$ $2.7 \times 10^{2}$ \\ 
  & & & & & B & $5.9 \times 10^{7}$ $\pm$ $5.9 \times 10^{4}$ & $3.1 \times 10^{7}$ $\pm$ $4.3 \times 10^{4}$ & $7.2 \times 10^{4}$ $\pm$ $2.1 \times 10^{3}$ \\ 
 \hline
 \multirow{2}{*}{10} & \multirow{2}{*}{\textless 2.1} & \multirow{2}{*}{\textless 0.6} & \multirow{2}{*}{\textless 1.6} & \multirow{2}{*}{\textless 0.4} & S & $4.1 \times 10^{5}$ $\pm$ $4.1 \times 10^{2}$ & $3.6 \times 10^{5}$ $\pm$ $3.8 \times 10^{2}$ & $1.9 \times 10^{5}$ $\pm$ $2.8 \times 10^{2}$ \\ 
  & & & & & B & $5.9 \times 10^{7}$ $\pm$ $5.9 \times 10^{4}$ & $3.1 \times 10^{7}$ $\pm$ $4.3 \times 10^{4}$ & $1.3 \times 10^{5}$ $\pm$ $2.8 \times 10^{3}$ \\ 
 \hline
 \multirow{2}{*}{25} & \multirow{2}{*}{\textless 2.0} & \multirow{2}{*}{\textless 0.7} & \multirow{2}{*}{\textless 1.6} & \multirow{2}{*}{\textless 0.8} & S & $3.8 \times 10^{5}$ $\pm$ $3.9 \times 10^{2}$ & $3.4 \times 10^{5}$ $\pm$ $3.7 \times 10^{2}$ & $1.9 \times 10^{5}$ $\pm$ $2.7 \times 10^{2}$ \\ 
  & & & & & B & $5.9 \times 10^{7}$ $\pm$ $5.9 \times 10^{4}$ & $3.1 \times 10^{7}$ $\pm$ $4.3 \times 10^{4}$ & $2.0 \times 10^{5}$ $\pm$ $3.5 \times 10^{3}$ \\ 
 \hline
 \multirow{2}{*}{50} & \multirow{2}{*}{\textless 2.0} & \multirow{2}{*}{\textless 0.8} & \multirow{2}{*}{\textless 1.3} & \multirow{2}{*}{\textless 1.7} & S & $3.0 \times 10^{5}$ $\pm$ $3.0 \times 10^{2}$ & $2.7 \times 10^{5}$ $\pm$ $2.9 \times 10^{2}$ & $1.6 \times 10^{5}$ $\pm$ $2.2 \times 10^{2}$ \\ 
  & & & & & B & $5.9 \times 10^{7}$ $\pm$ $5.9 \times 10^{4}$ & $3.1 \times 10^{7}$ $\pm$ $4.3 \times 10^{4}$ & $1.6 \times 10^{5}$ $\pm$ $3.0 \times 10^{3}$ \\ 
 \hline
 \multirow{2}{*}{60} & \multirow{2}{*}{\textless 2.0} & \multirow{2}{*}{\textless 0.8} & \multirow{2}{*}{\textless 1.2} & \multirow{2}{*}{\textless 1.9} & S & $2.6 \times 10^{5}$ $\pm$ $2.6 \times 10^{2}$ & $2.4 \times 10^{5}$ $\pm$ $2.5 \times 10^{2}$ & $1.2 \times 10^{5}$ $\pm$ $1.8 \times 10^{2}$ \\ 
  & & & & & B & $5.9 \times 10^{7}$ $\pm$ $5.9 \times 10^{4}$ & $3.1 \times 10^{7}$ $\pm$ $4.3 \times 10^{4}$ & $1.1 \times 10^{5}$ $\pm$ $2.6 \times 10^{3}$ \\ 
 \hline
 \multirow{2}{*}{70} & \multirow{2}{*}{\textless 1.9} & \multirow{2}{*}{\textless 0.8} & \multirow{2}{*}{\textless 1.3} & \multirow{2}{*}{\textless 2.1} & S & $2.2 \times 10^{5}$ $\pm$ $2.2 \times 10^{2}$ & $2.0 \times 10^{5}$ $\pm$ $2.1 \times 10^{2}$ & $1.1 \times 10^{5}$ $\pm$ $1.6 \times 10^{2}$ \\ 
  & & & & & B & $5.9 \times 10^{7}$ $\pm$ $5.9 \times 10^{4}$ & $3.1 \times 10^{7}$ $\pm$ $4.3 \times 10^{4}$ & $1.1 \times 10^{5}$ $\pm$ $2.5 \times 10^{3}$ \\ 
 \hline
 \multirow{2}{*}{80} & \multirow{2}{*}{\textless 2.0} & \multirow{2}{*}{\textless 0.8} & \multirow{2}{*}{\textless 1.1} & \multirow{2}{*}{\textless 2.2} & S & $1.7 \times 10^{5}$ $\pm$ $1.8 \times 10^{2}$ & $1.6 \times 10^{5}$ $\pm$ $1.7 \times 10^{2}$ & $7.1 \times 10^{4}$ $\pm$ $1.1 \times 10^{2}$ \\ 
  & & & & & B & $5.9 \times 10^{7}$ $\pm$ $5.9 \times 10^{4}$ & $3.1 \times 10^{7}$ $\pm$ $4.3 \times 10^{4}$ & $6.2 \times 10^{4}$ $\pm$ $1.9 \times 10^{3}$ \\ 
 \hline
 \multirow{2}{*}{90} & \multirow{2}{*}{\textless 1.8} & \multirow{2}{*}{\textless 0.8} & \multirow{2}{*}{\textless 1.2} & \multirow{2}{*}{\textless 2.4} & S & $1.3 \times 10^{5}$ $\pm$ $1.3 \times 10^{2}$ & $1.2 \times 10^{5}$ $\pm$ $1.3 \times 10^{2}$ & $5.7 \times 10^{4}$ $\pm$ $8.7 \times 10^{1}$ \\ 
  & & & & & B & $5.9 \times 10^{7}$ $\pm$ $5.9 \times 10^{4}$ & $3.1 \times 10^{7}$ $\pm$ $4.3 \times 10^{4}$ & $6.2 \times 10^{4}$ $\pm$ $1.9 \times 10^{3}$ \\ 
 \hline
 \multirow{2}{*}{100} & \multirow{2}{*}{\textless 1.9} & \multirow{2}{*}{\textless 0.8} & \multirow{2}{*}{\textless 1.2} & \multirow{2}{*}{\textless 2.5} & S & $9.3 \times 10^{4}$ $\pm$ $9.4 \times 10^{1}$ & $8.4 \times 10^{4}$ $\pm$ $8.9 \times 10^{1}$ & $4.0 \times 10^{4}$ $\pm$ $6.2 \times 10^{1}$ \\ 
  & & & & & B & $5.9 \times 10^{7}$ $\pm$ $5.9 \times 10^{4}$ & $3.1 \times 10^{7}$ $\pm$ $4.3 \times 10^{4}$ & $5.7 \times 10^{4}$ $\pm$ $1.8 \times 10^{3}$ \\ 
 \hline
 \multirow{2}{*}{120} & \multirow{2}{*}{\textless 1.9} & \multirow{2}{*}{\textless 0.8} & \multirow{2}{*}{\textless 1.2} & \multirow{2}{*}{\textless 2.8} & S & $3.3 \times 10^{4}$ $\pm$ $3.3 \times 10^{1}$ & $3.0 \times 10^{4}$ $\pm$ $3.2 \times 10^{1}$ & $1.6 \times 10^{4}$ $\pm$ $2.3 \times 10^{1}$ \\ 
  & & & & & B & $5.9 \times 10^{7}$ $\pm$ $5.9 \times 10^{4}$ & $3.1 \times 10^{7}$ $\pm$ $4.3 \times 10^{4}$ & $7.0 \times 10^{4}$ $\pm$ $2.0 \times 10^{3}$ \\ 
 \hline
 \multirow{2}{*}{140} & \multirow{2}{*}{\textless 2.0} & \multirow{2}{*}{\textless 0.9} & \multirow{2}{*}{\textless 1.2} & \multirow{2}{*}{\textless 3.0} & S & $4.8 \times 10^{3}$ $\pm$ $4.8 \times 10^{0}$ & $4.3 \times 10^{3}$ $\pm$ $4.6 \times 10^{0}$ & $2.5 \times 10^{3}$ $\pm$ $3.5 \times 10^{0}$ \\ 
  & & & & & B & $5.9 \times 10^{7}$ $\pm$ $5.9 \times 10^{4}$ & $3.1 \times 10^{7}$ $\pm$ $4.3 \times 10^{4}$ & $1.0 \times 10^{5}$ $\pm$ $2.5 \times 10^{3}$ \\ 
 \hline
 \multirow{2}{*}{150} & \multirow{2}{*}{\textless 1.8} & \multirow{2}{*}{\textless 0.8} & \multirow{2}{*}{\textless 1.2} & \multirow{2}{*}{\textless 3.1} & S & $6.4 \times 10^{2}$ $\pm$ $6.5 \times 10^{-1}$ & $5.8 \times 10^{2}$ $\pm$ $6.2 \times 10^{-1}$ & $2.9 \times 10^{2}$ $\pm$ $4.4 \times 10^{-1}$ \\ 
  & & & & & B & $5.9 \times 10^{7}$ $\pm$ $5.9 \times 10^{4}$ & $3.1 \times 10^{7}$ $\pm$ $4.3 \times 10^{4}$ & $9.7 \times 10^{4}$ $\pm$ $2.4 \times 10^{3}$ \\ 
\end{tabular}
}
\caption{Event selections at the WW running stage, for each mass point. The rightmost columns show the number of events for no applied selection, after the preselection criteria, and after the final selection criteria, for both signal (S) and background (B).}
\label{ww_selections}
\end{center}
\end{table*}

\begin{table*}[tbp]
\begin{center}
\resizebox{\linewidth}{!}{
\begin{tabular}{ c || c | c | c | c || c | c | c | c }
\multicolumn{9}{c}{ZH running stage} \\
\hline\hline
 $m_a$ & \multirow{2}{*}{$M_{cut}$} & \multirow{2}{*}{$|\cos{\theta_{\gamma_1}}|$} & \multirow{2}{*}{$|\phi_{\gamma_1}|$} & \multirow{2}{*}{$\Delta \alpha_{\gamma_1 \gamma_2}$} & \multirow{2}{*}{} & \multirow{2}{*}{No selection} & \multirow{2}{*}{Preselection} & \multirow{2}{*}{Final selection} \\ [0.0ex]
 [GeV] & & & & & & & & \\ 
 \hline 
 \multirow{2}{*}{5} & \multirow{2}{*}{\textless 2.1} & \multirow{2}{*}{\textless 0.6} & \multirow{2}{*}{\textless 1.6} & \multirow{2}{*}{\textless 0.2} & S & $2.2 \times 10^{5}$ $\pm$ $2.2 \times 10^{2}$ & $2.0 \times 10^{5}$ $\pm$ $2.1 \times 10^{2}$ & $1.0 \times 10^{5}$ $\pm$ $1.5 \times 10^{2}$ \\ 
  & & & & & B & $1.5 \times 10^{7}$ $\pm$ $1.5 \times 10^{4}$ & $8.4 \times 10^{6}$ $\pm$ $1.1 \times 10^{4}$ & $1.5 \times 10^{4}$ $\pm$ $4.8 \times 10^{2}$ \\ 
 \hline
 \multirow{2}{*}{10} & \multirow{2}{*}{\textless 2.1} & \multirow{2}{*}{\textless 0.6} & \multirow{2}{*}{\textless 1.6} & \multirow{2}{*}{\textless 0.3} & S & $2.2 \times 10^{5}$ $\pm$ $2.2 \times 10^{2}$ & $1.9 \times 10^{5}$ $\pm$ $2.1 \times 10^{2}$ & $1.0 \times 10^{5}$ $\pm$ $1.5 \times 10^{2}$ \\ 
  & & & & & B & $1.5 \times 10^{7}$ $\pm$ $1.5 \times 10^{4}$ & $8.4 \times 10^{6}$ $\pm$ $1.1 \times 10^{4}$ & $2.6 \times 10^{4}$ $\pm$ $6.3 \times 10^{2}$ \\ 
 \hline
 \multirow{2}{*}{25} & \multirow{2}{*}{\textless 2.1} & \multirow{2}{*}{\textless 0.6} & \multirow{2}{*}{\textless 1.6} & \multirow{2}{*}{\textless 0.6} & S & $1.9 \times 10^{5}$ $\pm$ $2.0 \times 10^{2}$ & $1.9 \times 10^{5}$ $\pm$ $2.0 \times 10^{2}$ & $9.7 \times 10^{4}$ $\pm$ $1.4 \times 10^{2}$ \\ 
  & & & & & B & $1.5 \times 10^{7}$ $\pm$ $1.5 \times 10^{4}$ & $8.4 \times 10^{6}$ $\pm$ $1.1 \times 10^{4}$ & $4.3 \times 10^{4}$ $\pm$ $8.1 \times 10^{2}$ \\ 
 \hline
 \multirow{2}{*}{50} & \multirow{2}{*}{\textless 2.0} & \multirow{2}{*}{\textless 0.8} & \multirow{2}{*}{\textless 1.6} & \multirow{2}{*}{\textless 1.2} & S & $1.9 \times 10^{5}$ $\pm$ $1.9 \times 10^{2}$ & $1.7 \times 10^{5}$ $\pm$ $1.8 \times 10^{2}$ & $1.1 \times 10^{5}$ $\pm$ $1.5 \times 10^{2}$ \\ 
  & & & & & B & $1.5 \times 10^{7}$ $\pm$ $1.5 \times 10^{4}$ & $8.4 \times 10^{6}$ $\pm$ $1.1 \times 10^{4}$ & $6.4 \times 10^{4}$ $\pm$ $9.9 \times 10^{2}$ \\ 
 \hline
 \multirow{2}{*}{60} & \multirow{2}{*}{\textless 2.1} & \multirow{2}{*}{\textless 0.8} & \multirow{2}{*}{\textless 1.6} & \multirow{2}{*}{\textless 1.4} & S & $1.8 \times 10^{5}$ $\pm$ $1.8 \times 10^{2}$ & $1.6 \times 10^{5}$ $\pm$ $1.7 \times 10^{2}$ & $1.1 \times 10^{5}$ $\pm$ $1.4 \times 10^{2}$ \\ 
  & & & & & B & $1.5 \times 10^{7}$ $\pm$ $1.5 \times 10^{4}$ & $8.4 \times 10^{6}$ $\pm$ $1.1 \times 10^{4}$ & $6.4 \times 10^{4}$ $\pm$ $9.9 \times 10^{2}$ \\ 
 \hline
 \multirow{2}{*}{70} & \multirow{2}{*}{\textless 2.0} & \multirow{2}{*}{\textless 0.9} & \multirow{2}{*}{\textless 1.6} & \multirow{2}{*}{\textless 1.7} & S & $1.7 \times 10^{5}$ $\pm$ $1.7 \times 10^{2}$ & $1.5 \times 10^{5}$ $\pm$ $1.6 \times 10^{2}$ & $1.1 \times 10^{5}$ $\pm$ $1.3 \times 10^{2}$ \\ 
  & & & & & B & $1.5 \times 10^{7}$ $\pm$ $1.5 \times 10^{4}$ & $8.4 \times 10^{6}$ $\pm$ $1.1 \times 10^{4}$ & $6.2 \times 10^{4}$ $\pm$ $9.7 \times 10^{2}$ \\ 
 \hline
 \multirow{2}{*}{80} & \multirow{2}{*}{\textless 1.9} & \multirow{2}{*}{\textless 0.8} & \multirow{2}{*}{\textless 1.6} & \multirow{2}{*}{\textless 1.7} & S & $1.5 \times 10^{5}$ $\pm$ $1.5 \times 10^{2}$ & $1.4 \times 10^{5}$ $\pm$ $1.5 \times 10^{2}$ & $8.6 \times 10^{4}$ $\pm$ $1.2 \times 10^{2}$ \\ 
  & & & & & B & $1.5 \times 10^{7}$ $\pm$ $1.5 \times 10^{4}$ & $8.4 \times 10^{6}$ $\pm$ $1.1 \times 10^{4}$ & $4.3 \times 10^{4}$ $\pm$ $8.2 \times 10^{2}$ \\ 
 \hline
 \multirow{2}{*}{90} & \multirow{2}{*}{\textless 2.0} & \multirow{2}{*}{\textless 0.8} & \multirow{2}{*}{\textless 1.6} & \multirow{2}{*}{\textless 1.9} & S & $1.4 \times 10^{5}$ $\pm$ $1.4 \times 10^{2}$ & $1.2 \times 10^{5}$ $\pm$ $1.3 \times 10^{2}$ & $8.4 \times 10^{4}$ $\pm$ $1.1 \times 10^{2}$ \\ 
  & & & & & B & $1.5 \times 10^{7}$ $\pm$ $1.5 \times 10^{4}$ & $8.4 \times 10^{6}$ $\pm$ $1.1 \times 10^{4}$ & $4.4 \times 10^{4}$ $\pm$ $8.3 \times 10^{2}$ \\ 
 \hline
 \multirow{2}{*}{100} & \multirow{2}{*}{\textless 1.9} & \multirow{2}{*}{\textless 0.8} & \multirow{2}{*}{\textless 1.6} & \multirow{2}{*}{\textless 2.1} & S & $1.2 \times 10^{5}$ $\pm$ $1.2 \times 10^{2}$ & $1.1 \times 10^{5}$ $\pm$ $1.2 \times 10^{2}$ & $7.4 \times 10^{4}$ $\pm$ $9.6 \times 10^{1}$ \\ 
  & & & & & B & $1.5 \times 10^{7}$ $\pm$ $1.5 \times 10^{4}$ & $8.4 \times 10^{6}$ $\pm$ $1.1 \times 10^{4}$ & $4.2 \times 10^{4}$ $\pm$ $8.0 \times 10^{2}$ \\ 
 \hline
 \multirow{2}{*}{120} & \multirow{2}{*}{\textless 1.9} & \multirow{2}{*}{\textless 0.9} & \multirow{2}{*}{\textless 1.6} & \multirow{2}{*}{\textless 2.3} & S & $8.8 \times 10^{4}$ $\pm$ $8.9 \times 10^{1}$ & $8.0 \times 10^{4}$ $\pm$ $8.4 \times 10^{1}$ & $5.4 \times 10^{4}$ $\pm$ $6.9 \times 10^{1}$ \\ 
  & & & & & B & $1.5 \times 10^{7}$ $\pm$ $1.5 \times 10^{4}$ & $8.4 \times 10^{6}$ $\pm$ $1.1 \times 10^{4}$ & $3.5 \times 10^{4}$ $\pm$ $7.4 \times 10^{2}$ \\ 
 \hline
 \multirow{2}{*}{140} & \multirow{2}{*}{\textless 2.0} & \multirow{2}{*}{\textless 0.8} & \multirow{2}{*}{\textless 1.6} & \multirow{2}{*}{\textless 2.4} & S & $5.2 \times 10^{4}$ $\pm$ $5.4 \times 10^{1}$ & $5.1 \times 10^{4}$ $\pm$ $5.4 \times 10^{1}$ & $3.2 \times 10^{4}$ $\pm$ $4.3 \times 10^{1}$ \\ 
  & & & & & B & $1.5 \times 10^{7}$ $\pm$ $1.5 \times 10^{4}$ & $8.4 \times 10^{6}$ $\pm$ $1.1 \times 10^{4}$ & $2.8 \times 10^{4}$ $\pm$ $6.6 \times 10^{2}$ \\ 
 \hline
 \multirow{2}{*}{150} & \multirow{2}{*}{\textless 1.9} & \multirow{2}{*}{\textless 0.8} & \multirow{2}{*}{\textless 1.6} & \multirow{2}{*}{\textless 2.6} & S & $4.3 \times 10^{4}$ $\pm$ $4.4 \times 10^{1}$ & $3.9 \times 10^{4}$ $\pm$ $4.1 \times 10^{1}$ & $2.6 \times 10^{4}$ $\pm$ $3.4 \times 10^{1}$ \\ 
  & & & & & B & $1.5 \times 10^{7}$ $\pm$ $1.5 \times 10^{4}$ & $8.4 \times 10^{6}$ $\pm$ $1.1 \times 10^{4}$ & $3.0 \times 10^{4}$ $\pm$ $6.8 \times 10^{2}$ \\ 
 \hline
 \multirow{2}{*}{160} & \multirow{2}{*}{\textless 2.0} & \multirow{2}{*}{\textless 0.9} & \multirow{2}{*}{\textless 1.6} & \multirow{2}{*}{\textless 2.7} & S & $3.2 \times 10^{4}$ $\pm$ $3.2 \times 10^{1}$ & $2.9 \times 10^{4}$ $\pm$ $3.1 \times 10^{1}$ & $2.1 \times 10^{4}$ $\pm$ $2.6 \times 10^{1}$ \\ 
  & & & & & B & $1.5 \times 10^{7}$ $\pm$ $1.5 \times 10^{4}$ & $8.4 \times 10^{6}$ $\pm$ $1.1 \times 10^{4}$ & $3.5 \times 10^{4}$ $\pm$ $7.3 \times 10^{2}$ \\ 
 \hline
 \multirow{2}{*}{180} & \multirow{2}{*}{\textless 2.0} & \multirow{2}{*}{\textless 1.0} & \multirow{2}{*}{\textless 1.6} & \multirow{2}{*}{\textless 2.9} & S & $1.5 \times 10^{4}$ $\pm$ $1.5 \times 10^{1}$ & $1.3 \times 10^{4}$ $\pm$ $1.4 \times 10^{1}$ & $1.1 \times 10^{4}$ $\pm$ $1.3 \times 10^{1}$ \\ 
  & & & & & B & $1.5 \times 10^{7}$ $\pm$ $1.5 \times 10^{4}$ & $8.4 \times 10^{6}$ $\pm$ $1.1 \times 10^{4}$ & $4.1 \times 10^{4}$ $\pm$ $7.9 \times 10^{2}$ \\ 
 \hline
 \multirow{2}{*}{200} & \multirow{2}{*}{\textless 1.9} & \multirow{2}{*}{\textless 0.9} & \multirow{2}{*}{\textless 1.6} & \multirow{2}{*}{\textless 3.0} & S & $4.9 \times 10^{3}$ $\pm$ $4.9 \times 10^{0}$ & $4.4 \times 10^{3}$ $\pm$ $4.6 \times 10^{0}$ & $3.3 \times 10^{3}$ $\pm$ $4.0 \times 10^{0}$ \\ 
  & & & & & B & $1.5 \times 10^{7}$ $\pm$ $1.5 \times 10^{4}$ & $8.4 \times 10^{6}$ $\pm$ $1.1 \times 10^{4}$ & $3.8 \times 10^{4}$ $\pm$ $7.7 \times 10^{2}$ \\ 
 \hline
 \multirow{2}{*}{220} & \multirow{2}{*}{\textless 1.9} & \multirow{2}{*}{\textless 0.9} & \multirow{2}{*}{\textless 1.6} & \multirow{2}{*}{\textless 3.1} & S & $6.7 \times 10^{2}$ $\pm$ $6.7 \times 10^{-1}$ & $6.0 \times 10^{2}$ $\pm$ $6.4 \times 10^{-1}$ & $4.6 \times 10^{2}$ $\pm$ $5.5 \times 10^{-1}$ \\ 
  & & & & & B & $1.5 \times 10^{7}$ $\pm$ $1.5 \times 10^{4}$ & $8.4 \times 10^{6}$ $\pm$ $1.1 \times 10^{4}$ & $4.7 \times 10^{4}$ $\pm$ $8.5 \times 10^{2}$ \\ 
\end{tabular}
}
\caption{Event selections at the ZH running stage, for each mass point. The rightmost columns show the number of events for no applied selection, after the preselection criteria, and after the final selection criteria, for both signal (S) and background (B).}
\label{zh_selections}
\end{center}
\end{table*}

\begin{table*}[tbp]
\begin{center}
\resizebox{\linewidth}{!}{
\begin{tabular}{ c || c | c | c | c || c | c | c | c }
\multicolumn{9}{c}{\ttbar running stage} \\
\hline\hline
 $m_a$ & \multirow{2}{*}{$M_{cut}$} & \multirow{2}{*}{$|\cos{\theta_{\gamma_1}}|$} & \multirow{2}{*}{$|\phi_{\gamma_1}|$} & \multirow{2}{*}{$\Delta \alpha_{\gamma_1 \gamma_2}$} & \multirow{2}{*}{} & \multirow{2}{*}{No selection} & \multirow{2}{*}{Preselection} & \multirow{2}{*}{Final selection} \\ [0.0ex]
 [GeV] & & & & & & & & \\ 
 \hline 
 \multirow{2}{*}{5} & \multirow{2}{*}{\textless 2.1} & \multirow{2}{*}{\textless 0.6} & \multirow{2}{*}{\textless 1.6} & \multirow{2}{*}{\textless 0.1} & S & $5.4 \times 10^{4}$ $\pm$ $5.4 \times 10^{1}$ & $4.9 \times 10^{4}$ $\pm$ $5.2 \times 10^{1}$ & $2.6 \times 10^{4}$ $\pm$ $3.7 \times 10^{1}$ \\ 
  & & & & & B & $1.8 \times 10^{6}$ $\pm$ $1.8 \times 10^{3}$ & $1.0 \times 10^{6}$ $\pm$ $1.4 \times 10^{3}$ & $1.2 \times 10^{3}$ $\pm$ $4.7 \times 10^{1}$ \\ 
 \hline
 \multirow{2}{*}{10} & \multirow{2}{*}{\textless 2.1} & \multirow{2}{*}{\textless 0.7} & \multirow{2}{*}{\textless 1.6} & \multirow{2}{*}{\textless 0.2} & S & $5.4 \times 10^{4}$ $\pm$ $5.4 \times 10^{1}$ & $4.9 \times 10^{4}$ $\pm$ $5.1 \times 10^{1}$ & $3.0 \times 10^{4}$ $\pm$ $4.0 \times 10^{1}$ \\ 
  & & & & & B & $1.8 \times 10^{6}$ $\pm$ $1.8 \times 10^{3}$ & $1.0 \times 10^{6}$ $\pm$ $1.4 \times 10^{3}$ & $2.9 \times 10^{3}$ $\pm$ $7.2 \times 10^{1}$ \\ 
 \hline
 \multirow{2}{*}{25} & \multirow{2}{*}{\textless 2.2} & \multirow{2}{*}{\textless 0.6} & \multirow{2}{*}{\textless 1.6} & \multirow{2}{*}{\textless 0.4} & S & $5.3 \times 10^{4}$ $\pm$ $5.3 \times 10^{1}$ & $4.7 \times 10^{4}$ $\pm$ $5.0 \times 10^{1}$ & $2.5 \times 10^{4}$ $\pm$ $3.6 \times 10^{1}$ \\ 
  & & & & & B & $1.8 \times 10^{6}$ $\pm$ $1.8 \times 10^{3}$ & $1.0 \times 10^{6}$ $\pm$ $1.4 \times 10^{3}$ & $4.0 \times 10^{3}$ $\pm$ $8.4 \times 10^{1}$ \\ 
 \hline
 \multirow{2}{*}{50} & \multirow{2}{*}{\textless 2.0} & \multirow{2}{*}{\textless 0.7} & \multirow{2}{*}{\textless 1.6} & \multirow{2}{*}{\textless 0.8} & S & $5.1 \times 10^{4}$ $\pm$ $5.1 \times 10^{1}$ & $4.6 \times 10^{4}$ $\pm$ $4.8 \times 10^{1}$ & $2.6 \times 10^{4}$ $\pm$ $3.7 \times 10^{1}$ \\ 
  & & & & & B & $1.8 \times 10^{6}$ $\pm$ $1.8 \times 10^{3}$ & $1.0 \times 10^{6}$ $\pm$ $1.4 \times 10^{3}$ & $6.3 \times 10^{3}$ $\pm$ $1.1 \times 10^{2}$ \\ 
 \hline
 \multirow{2}{*}{60} & \multirow{2}{*}{\textless 1.8} & \multirow{2}{*}{\textless 0.8} & \multirow{2}{*}{\textless 1.6} & \multirow{2}{*}{\textless 0.9} & S & $4.9 \times 10^{4}$ $\pm$ $5.0 \times 10^{1}$ & $4.4 \times 10^{4}$ $\pm$ $4.7 \times 10^{1}$ & $2.6 \times 10^{4}$ $\pm$ $3.6 \times 10^{1}$ \\ 
  & & & & & B & $1.8 \times 10^{6}$ $\pm$ $1.8 \times 10^{3}$ & $1.0 \times 10^{6}$ $\pm$ $1.4 \times 10^{3}$ & $5.7 \times 10^{3}$ $\pm$ $1.0 \times 10^{2}$ \\ 
 \hline
 \multirow{2}{*}{70} & \multirow{2}{*}{\textless 1.9} & \multirow{2}{*}{\textless 0.8} & \multirow{2}{*}{\textless 1.6} & \multirow{2}{*}{\textless 1.1} & S & $4.8 \times 10^{4}$ $\pm$ $4.8 \times 10^{1}$ & $4.3 \times 10^{4}$ $\pm$ $4.6 \times 10^{1}$ & $2.7 \times 10^{4}$ $\pm$ $3.6 \times 10^{1}$ \\ 
  & & & & & B & $1.8 \times 10^{6}$ $\pm$ $1.8 \times 10^{3}$ & $1.0 \times 10^{6}$ $\pm$ $1.4 \times 10^{3}$ & $6.4 \times 10^{3}$ $\pm$ $1.1 \times 10^{2}$ \\ 
 \hline
 \multirow{2}{*}{80} & \multirow{2}{*}{\textless 1.9} & \multirow{2}{*}{\textless 0.8} & \multirow{2}{*}{\textless 1.6} & \multirow{2}{*}{\textless 1.3} & S & $4.6 \times 10^{4}$ $\pm$ $4.7 \times 10^{1}$ & $4.2 \times 10^{4}$ $\pm$ $4.4 \times 10^{1}$ & $2.7 \times 10^{4}$ $\pm$ $3.6 \times 10^{1}$ \\ 
  & & & & & B & $1.8 \times 10^{6}$ $\pm$ $1.8 \times 10^{3}$ & $1.0 \times 10^{6}$ $\pm$ $1.4 \times 10^{3}$ & $6.5 \times 10^{3}$ $\pm$ $1.1 \times 10^{2}$ \\ 
 \hline
 \multirow{2}{*}{90} & \multirow{2}{*}{\textless 2.0} & \multirow{2}{*}{\textless 0.8} & \multirow{2}{*}{\textless 1.6} & \multirow{2}{*}{\textless 1.3} & S & $4.4 \times 10^{4}$ $\pm$ $4.5 \times 10^{1}$ & $4.0 \times 10^{4}$ $\pm$ $4.2 \times 10^{1}$ & $2.4 \times 10^{4}$ $\pm$ $3.3 \times 10^{1}$ \\ 
  & & & & & B & $1.8 \times 10^{6}$ $\pm$ $1.8 \times 10^{3}$ & $1.0 \times 10^{6}$ $\pm$ $1.4 \times 10^{3}$ & $5.5 \times 10^{3}$ $\pm$ $9.9 \times 10^{1}$ \\ 
 \hline
 \multirow{2}{*}{100} & \multirow{2}{*}{\textless 1.9} & \multirow{2}{*}{\textless 0.8} & \multirow{2}{*}{\textless 1.6} & \multirow{2}{*}{\textless 1.6} & S & $4.2 \times 10^{4}$ $\pm$ $4.3 \times 10^{1}$ & $3.8 \times 10^{4}$ $\pm$ $4.0 \times 10^{1}$ & $2.5 \times 10^{4}$ $\pm$ $3.3 \times 10^{1}$ \\ 
  & & & & & B & $1.8 \times 10^{6}$ $\pm$ $1.8 \times 10^{3}$ & $1.0 \times 10^{6}$ $\pm$ $1.4 \times 10^{3}$ & $5.9 \times 10^{3}$ $\pm$ $1.0 \times 10^{2}$ \\ 
 \hline
 \multirow{2}{*}{120} & \multirow{2}{*}{\textless 1.9} & \multirow{2}{*}{\textless 0.9} & \multirow{2}{*}{\textless 1.6} & \multirow{2}{*}{\textless 1.8} & S & $3.7 \times 10^{4}$ $\pm$ $3.7 \times 10^{1}$ & $3.3 \times 10^{4}$ $\pm$ $3.5 \times 10^{1}$ & $2.2 \times 10^{4}$ $\pm$ $2.9 \times 10^{1}$ \\ 
  & & & & & B & $1.8 \times 10^{6}$ $\pm$ $1.8 \times 10^{3}$ & $1.0 \times 10^{6}$ $\pm$ $1.4 \times 10^{3}$ & $5.3 \times 10^{3}$ $\pm$ $9.7 \times 10^{1}$ \\ 
 \hline
 \multirow{2}{*}{140} & \multirow{2}{*}{\textless 2.0} & \multirow{2}{*}{\textless 0.8} & \multirow{2}{*}{\textless 1.6} & \multirow{2}{*}{\textless 2.0} & S & $3.0 \times 10^{4}$ $\pm$ $3.0 \times 10^{1}$ & $2.7 \times 10^{4}$ $\pm$ $2.9 \times 10^{1}$ & $1.9 \times 10^{4}$ $\pm$ $2.4 \times 10^{1}$ \\ 
  & & & & & B & $1.8 \times 10^{6}$ $\pm$ $1.8 \times 10^{3}$ & $1.0 \times 10^{6}$ $\pm$ $1.4 \times 10^{3}$ & $4.9 \times 10^{3}$ $\pm$ $9.4 \times 10^{1}$ \\ 
 \hline
 \multirow{2}{*}{150} & \multirow{2}{*}{\textless 2.1} & \multirow{2}{*}{\textless 0.8} & \multirow{2}{*}{\textless 1.3} & \multirow{2}{*}{\textless 2.0} & S & $2.7 \times 10^{4}$ $\pm$ $2.7 \times 10^{1}$ & $2.4 \times 10^{4}$ $\pm$ $2.6 \times 10^{1}$ & $1.4 \times 10^{4}$ $\pm$ $2.0 \times 10^{1}$ \\ 
  & & & & & B & $1.8 \times 10^{6}$ $\pm$ $1.8 \times 10^{3}$ & $1.0 \times 10^{6}$ $\pm$ $1.4 \times 10^{3}$ & $3.1 \times 10^{3}$ $\pm$ $7.5 \times 10^{1}$ \\ 
 \hline
 \multirow{2}{*}{160} & \multirow{2}{*}{\textless 2.0} & \multirow{2}{*}{\textless 0.8} & \multirow{2}{*}{\textless 1.2} & \multirow{2}{*}{\textless 2.2} & S & $2.4 \times 10^{4}$ $\pm$ $2.4 \times 10^{1}$ & $2.2 \times 10^{4}$ $\pm$ $2.3 \times 10^{1}$ & $1.2 \times 10^{4}$ $\pm$ $1.7 \times 10^{1}$ \\ 
  & & & & & B & $1.8 \times 10^{6}$ $\pm$ $1.8 \times 10^{3}$ & $1.0 \times 10^{6}$ $\pm$ $1.4 \times 10^{3}$ & $2.5 \times 10^{3}$ $\pm$ $6.7 \times 10^{1}$ \\ 
 \hline
 \multirow{2}{*}{180} & \multirow{2}{*}{\textless 2.1} & \multirow{2}{*}{\textless 0.8} & \multirow{2}{*}{\textless 1.3} & \multirow{2}{*}{\textless 2.2} & S & $1.9 \times 10^{4}$ $\pm$ $1.9 \times 10^{1}$ & $1.7 \times 10^{4}$ $\pm$ $1.8 \times 10^{1}$ & $9.6 \times 10^{3}$ $\pm$ $1.4 \times 10^{1}$ \\ 
  & & & & & B & $1.8 \times 10^{6}$ $\pm$ $1.8 \times 10^{3}$ & $1.0 \times 10^{6}$ $\pm$ $1.4 \times 10^{3}$ & $2.5 \times 10^{3}$ $\pm$ $6.8 \times 10^{1}$ \\ 
 \hline
 \multirow{2}{*}{200} & \multirow{2}{*}{\textless 2.0} & \multirow{2}{*}{\textless 0.8} & \multirow{2}{*}{\textless 1.2} & \multirow{2}{*}{\textless 2.3} & S & $1.4 \times 10^{4}$ $\pm$ $1.4 \times 10^{1}$ & $1.3 \times 10^{4}$ $\pm$ $1.4 \times 10^{1}$ & $6.2 \times 10^{3}$ $\pm$ $9.5 \times 10^{0}$ \\ 
  & & & & & B & $1.8 \times 10^{6}$ $\pm$ $1.8 \times 10^{3}$ & $1.0 \times 10^{6}$ $\pm$ $1.4 \times 10^{3}$ & $1.5 \times 10^{3}$ $\pm$ $5.2 \times 10^{1}$ \\ 
 \hline
 \multirow{2}{*}{220} & \multirow{2}{*}{\textless 2.0} & \multirow{2}{*}{\textless 0.8} & \multirow{2}{*}{\textless 1.3} & \multirow{2}{*}{\textless 2.5} & S & $1.0 \times 10^{4}$ $\pm$ $1.0 \times 10^{1}$ & $9.3 \times 10^{3}$ $\pm$ $9.9 \times 10^{0}$ & $5.3 \times 10^{3}$ $\pm$ $7.5 \times 10^{0}$ \\ 
  & & & & & B & $1.8 \times 10^{6}$ $\pm$ $1.8 \times 10^{3}$ & $1.0 \times 10^{6}$ $\pm$ $1.4 \times 10^{3}$ & $2.0 \times 10^{3}$ $\pm$ $6.0 \times 10^{1}$ \\ 
 \hline
 \multirow{2}{*}{240} & \multirow{2}{*}{\textless 1.9} & \multirow{2}{*}{\textless 0.8} & \multirow{2}{*}{\textless 1.2} & \multirow{2}{*}{\textless 2.6} & S & $7.1 \times 10^{3}$ $\pm$ $7.2 \times 10^{0}$ & $6.4 \times 10^{3}$ $\pm$ $6.8 \times 10^{0}$ & $3.2 \times 10^{3}$ $\pm$ $4.8 \times 10^{0}$ \\ 
  & & & & & B & $1.8 \times 10^{6}$ $\pm$ $1.8 \times 10^{3}$ & $1.0 \times 10^{6}$ $\pm$ $1.4 \times 10^{3}$ & $1.4 \times 10^{3}$ $\pm$ $5.0 \times 10^{1}$ \\ 
 \hline
 \multirow{2}{*}{300} & \multirow{2}{*}{\textless 2.0} & \multirow{2}{*}{\textless 0.8} & \multirow{2}{*}{\textless 1.3} & \multirow{2}{*}{\textless 3.0} & S & $1.2 \times 10^{3}$ $\pm$ $1.3 \times 10^{0}$ & $1.1 \times 10^{3}$ $\pm$ $1.2 \times 10^{0}$ & $6.6 \times 10^{2}$ $\pm$ $9.1 \times 10^{-1}$ \\ 
  & & & & & B & $1.8 \times 10^{6}$ $\pm$ $1.8 \times 10^{3}$ & $1.0 \times 10^{6}$ $\pm$ $1.4 \times 10^{3}$ & $2.1 \times 10^{3}$ $\pm$ $6.2 \times 10^{1}$ \\ 
 \hline
 \multirow{2}{*}{320} & \multirow{2}{*}{\textless 2.0} & \multirow{2}{*}{\textless 0.9} & \multirow{2}{*}{\textless 1.2} & \multirow{2}{*}{\textless 3.0} & S & $4.5 \times 10^{2}$ $\pm$ $4.5 \times 10^{-1}$ & $4.0 \times 10^{2}$ $\pm$ $4.3 \times 10^{-1}$ & $2.3 \times 10^{2}$ $\pm$ $3.2 \times 10^{-1}$ \\ 
  & & & & & B & $1.8 \times 10^{6}$ $\pm$ $1.8 \times 10^{3}$ & $1.0 \times 10^{6}$ $\pm$ $1.4 \times 10^{3}$ & $2.1 \times 10^{3}$ $\pm$ $6.1 \times 10^{1}$ \\ 
\end{tabular}
}
\caption{Event selections at the \ttbar running stage, for each mass point. The rightmost columns show the number of events for no applied selection, after the preselection criteria, and after the final selection criteria, for both signal (S) and background (B).}
\label{tt_selections}
\end{center}
\end{table*}

\end{appendices}

\clearpage
\enlargethispage{26\baselineskip}
\bibliography{sn-bibliography}

\end{document}